\documentclass[12pt,pra,reprint,superscriptaddress,amsmath,amsfonts,amssymb]{revtex4-2}

\usepackage[utf8]{inputenc}
\usepackage[T1]{fontenc}
\usepackage[sc,osf]{mathpazo}
\usepackage{amsmath, amsthm, amssymb,amsfonts,mathbbol,amstext}
\usepackage{graphicx}
\usepackage{dcolumn}
\usepackage{bm}
\usepackage{bbm}
\usepackage{mathtools}
\usepackage{comment}
\usepackage{multirow}
\usepackage{diagbox}
\usepackage{float}
\usepackage{tcolorbox}

\usepackage[pretty]{revquantum}
\usepackage[english]{babel}
\usepackage{graphicx}
\usepackage[caption=false]{subfig}
\usepackage{mathrsfs}
\usepackage{amsmath}
\usepackage{amsthm}
\usepackage{dsfont}
\usepackage{thmtools, thm-restate}
\usepackage{bm}
\usepackage{xcolor}
\usepackage{hyperref}
\usepackage{hypernat}
\usepackage[utf8]{inputenc}
\usepackage[T1]{fontenc}
\usepackage{cleveref}
\usepackage{soul}

\declaretheorem[parent=section,name=Theorem]{thm}

\declaretheorem[style=definition,sibling=thm]{definition}

\declaretheorem[sibling=thm]{proposition}
\declaretheorem[sibling=thm]{corollary}

\DeclareMathOperator*{\Minimise}{Minimise}
\DeclareMathOperator*{\Maximise}{Maximise}

\newcommand{\channel}[2]{\mathcal{N}_{#1|#2}}

\newcommand{\state}[2]{\varrho_{#1|#2}}
\newcommand{\ketbra}[2]{\ket{#1}\bra{#2}}
\newcommand{\dnorm}[1]{\|#1\|_\diamond}

\begin{document}
\title{Efficient and operational quantifier of non-divisibility in terms of channel discrimination}

\author{Ranieri Nery}
\affiliation{ICFO - Institut de Ciencies Fotoniques, The Barcelona Institute of Science and Technology, Castelldefels (Barcelona) 08860, Spain}
\author{Nadja K. Bernardes}
\affiliation{Departamento de Física, Centro de Ciências Exatas e da Natureza, Universidade Federal de Pernambuco, Recife-PE, 50670-901, Brazil}
\author{Daniel Cavalcanti}
\affiliation{Algorithmiq Ltd, Kanavakatu 3 C, FI-00160 Helsinki, Finland}
\author{Rafael Chaves}
\affiliation{International Institute of Physics, Federal University of Rio Grande do Norte, 59078-970, Natal, Brazil}
\affiliation{School of Science and Technology, Federal University of Rio Grande do Norte, Natal, Brazil}
\author{Cristhiano Duarte}
\affiliation{Fundação Maurício Grabois, R. Rego Freitas, 192 - República, São Paulo - SP, 01220-010, Brazil}
\affiliation{Instituto de Física, Universidade Federal da Bahia, Campus de Ondina, Rua Barão do Geremoabo, s.n., Ondina, Salvador, BA 40210-340, Brazil}
\affiliation{Institute for Quantum Studies, Chapman University, One University Drive, Orange, CA, 92866, USA}

\begin{abstract}
The understanding of open quantum systems is crucial for the development of quantum technologies. Of particular relevance is the characterisation of divisible quantum dynamics, seen as a generalisation of Markovian processes to the quantum setting. Here, we propose a way to detect divisibility and quantify how non-divisible a quantum channel is through the concept of channel discrimination. We ask how well we can distinguish generic dynamics from divisible dynamics. We show that this question can be answered efficiently through semidefinite programming, which provides us with an operational and efficient way to quantify non-divisibility.  
\end{abstract}

\maketitle

\section{Introduction}\label{Sec.Intro}

No physical system is completely isolated from its surrounding environment. The unavoidable interaction between a system and its environment renders the possibility of scaling quantum phenomena to the macroscopic world \cite{Zurek03,BKZ06,Schlosshauer05} and represents the main barrier for practical developments of new quantum technologies \cite{LCW98,AH14}. Unless one counteracts the detrimental effects of decoherence, which typically accumulate exponentially fast with the system's size \cite{Aolita_2008}, the advantages provided by quantum information processing with respect to classical strategies become unfeasible in practice. Because of this, a great effort is devoted to understanding and modeling decoherence processes and proposing methods that counteract their effect, such as noise mitigation \cite{Wallman_2016,Berg_2022, CaietAl23} and quantum error correction \cite{lidar2013quantum} schemes.

Formally, the system-environment interaction causes the system's evolution to be non-unitary, i.e., to be described by a more general, completely positive, and trace-preserving (CPTP) map. Given quantum dynamics, a natural question is whether it is Markovian or not \cite{RHP14,RevModPhys.88.021002}. Unlike the classical case, where Markovian concepts of memory effects and divisibility are linked, different notions of Markovianity arise in quantum mechanics \cite{CKR11,RHP14}. This has opened up an active field for discussions on how to properly define quantum Markovianity. Several proposals have defined Markovian processes, including channel divisibility \cite{RHP10}, non-increase of state distinguishability \cite{BLP,BD16,Bae_2016}, non-increase of system-environment correlations \cite{PhysRevA.96.062118,Dario1,Dario2,Janek,Abiuso_2022}, among others \cite{PhysRevLett.120.040405,PhysRevLett.121.240401, capela2020monogamy,capela2021quantum,PhysRevA.103.012218,PhysRevA.108.042202}. 

In this paper, we focus on the notion of divisibility and ask the following questions: (i) Can we efficiently detect whether a family of quantum channels is divisible? (ii) Given a non-divisible quantum dynamics, can we quantify its degree of non-divisibility in an efficient and meaningful operational way? We provide a positive answer to both questions by proposing quantifying non-divisibility as the minimum distance from the channel to a divisible one. We use a distance induced by the diamond norm between channels as a measure of distinguishability. Such distance can be calculated via semidefinite programming so that efficient numerical methods (i.e. polynomial in the dimension of the problem) can be used for its calculation~\cite{SC23}. Moreover, the distance induced by the diamond norm also has an operational interpretation regarding the minimum error one can make in a channel discrimination protocol when we use it for general probe states and measurements \cite{PirandolaEtAl19, BPH15}. We use this quantifier to investigate the amount of non-divisibility present in three different models. The first is a versatile toy model for studying open quantum system dynamics: the collisional model \cite{CICCARELLO20221}. The second is a paradigmatic decoherence channel: the dephasing model. Finally, the third model explores a convex combination of unitary channels. 

Note that in the context of a resource theory of Markovianity, channel discrimination and semidefinite programming for robustness have already been explored in Ref.~\cite{anand2019quantifying}. However, unlike the analysis presented here, the framework investigated in Ref.~\cite{anand2019quantifying} is only immediately applicable when the underlying set of Markovian dynamics forms a convex set. Moreover, the scheme proposed here is valid for all non-Markovian dynamics (invertible or not), which sets our work apart from universal non-Markovian quantifiers based on correlations \cite{Dario1,Dario2,Janek}. Similarly, recently we became aware of yet another novel measure of non-Markovianity, one that explores the relationship between CP indivisibility and the compatibility of quantum channels~\cite{MitraElAt24,DCD22}.

The paper is organized as follows. In Sec.~\ref{Sec.Divisibility} we review 
the concept of divisibility, P-divisibility, and CP-divisibility. In Sec.~\ref{Sec.CJIsomorphism}, the basic toolbox about conditional quantum states and the Choi–Jamiołkowski isomorphism are presented. In Sec.~\ref{Sec.SDPFormulation} we detail our main result, showing that checking whether a given dynamics is divisible (or not) can be cast as an efficient semi-definite program (SDP). In Sec.~\ref{SDPexample}, we apply our framework to characterize two examples of channels, in particular showing how witnesses and quantifiers of non-Markovianity can be extracted from it. In Sec.~\ref{Sec.Conclusion}, we discuss our findings while in Appendices \ref{App.ProofIsomorphism}, \ref{App.ProofCompositionLaw}, \ref{App.DiamondNormSDP} and \ref{App.AbsoluteNondiv} we include all the technical details required to understand and reproduce our results.

\section{Divisibility}\label{Sec.Divisibility}

\begin{figure}
    \centering
    \includegraphics[scale=0.26]{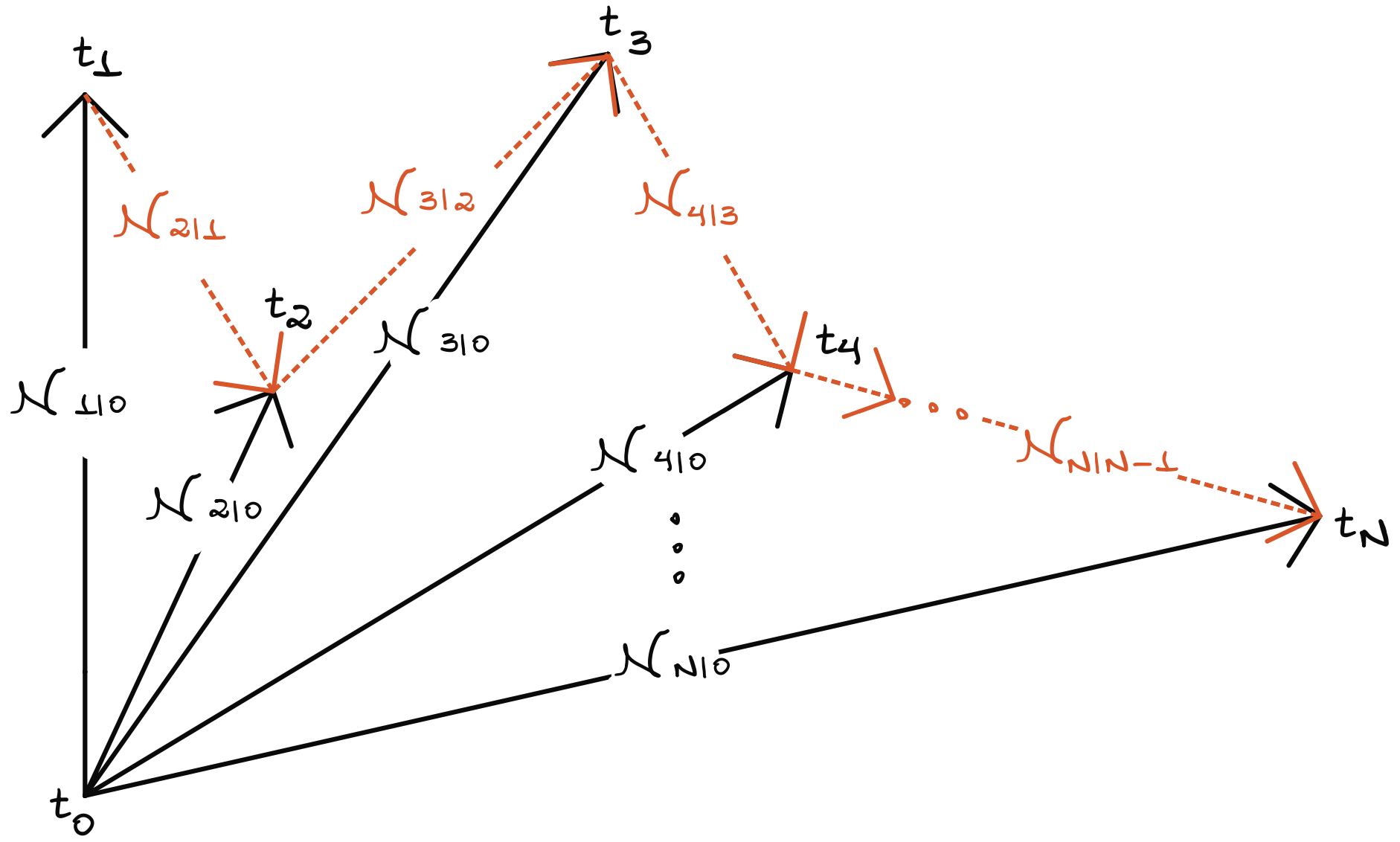}
    \caption{Schematic representation of discrete-time dynamics and the possibilities (dotted lines) for divisibility.} 
    \label{Fig.QuantumDynamics}
\end{figure}

The scenario we study in the paper is schematically depicted in Fig.~\ref{Fig.QuantumDynamics}. We investigate quantum dynamics that evolve quantum systems from time $t_0$ to time $t_N$. There is a paradigmatic way in which we can analyze these dynamics: by cutting the time interval $[t_{0},t_{N}]$ into $N$ steps and reconstructing (by means of quantum process tomography \cite{Mohseni_2008}) the maps that describe the time evolution from $t_0$ to each $t_k$ $(k=1,\cdots,N)$. Mathematically, each of these maps is described by a CPTP map 
\begin{equation}
\mathcal{N}_{k|0}:\mathcal{L}(\mathcal{H}_{0}) \rightarrow \mathcal{L}(\mathcal{H}_{k}), 
\end{equation}
where each $\mathcal{L}(\mathcal{H}_{k})$ represents the set of linear operators acting on the Hilbert space $\mathcal{H}_{k}$, and each $\mathcal{H}_{k}$ is associated with an instant of time. 
In other words, we are given a stroboscopic characterization of the dynamics in terms of the family of maps $\mathcal{F}:= \{\mathcal{N}_{k|0} \}_{k=1}^N$, each of which representing the time evolution of a quantum system from an initial time step $t_0$ to a given time step $t_k$. The fact that there is an apparent mismatch between the sub-index location ($k|0$) and the domain ($\mathcal{L}(\mathcal{H}_{0})$), contra-domain $(\mathcal{L}(\mathcal{H}_{k}))$ is explained in def.~\ref{Def.JamilIso} below. For now, the object $\mathcal{N}_{k | 0}$ should be read as the map governing the dynamics at $t_k$ given $t_0$. 

In this scenario, we say that a dynamics is \emph{divisible} whenever, given $\mathcal{F}$, we find a set of maps $ \{ \mathcal{N}_{k | k-1} \}_{k=1}^{N} $ such that 
\begin{align}\mathcal{N}_{k | 0} = \mathcal{N}_{k | k-1} \circ \mathcal{N}_{k -1 | 0}. \label{Eq.DefDivisibilityPairPair}
\end{align}
If no such decomposition can be found, the dynamics is said to be \emph{non-divisible}. Consequently, the divisibility of $\mathcal{F}$ imposes a net effect on $\mathcal{N}_{N | 0}$ in terms of the other intermediate maps:
\begin{equation}\label{divisibility}
    \mathcal{N}_{N|0}=\mathcal{N}_{N|N-1} \circ \cdots \circ \mathcal{N}_{2|1} \circ \mathcal{N}_{1|0}.
\end{equation}
In other words, we divide the map $\mathcal{N}_{N|0}$ into a sequence of maps $\mathcal{N}_{k | k-1}$. This situation is depicted in Fig.~\ref{Fig.QuantumDynamics}.

In the scenario characterised by eq.~\eqref{Eq.DefDivisibilityPairPair},  we can also consider two types of divisibility. We say that a dynamics is \emph{P-divisible} if we can find intermediate maps satisfying \eqref{Eq.DefDivisibilityPairPair} where the maps $\mathcal{N}_{k | k-1}$ are positive (but not necessarily completely positive), that is $\mathcal{N}_{k | k-1}(\rho)\geq 0,\, ~\forall ~\rho\geq0$. Usually, P-divisible dynamics that are also trace-preserving are dubbed PTP dynamics. Moreover, we say that a dynamics is \emph{CP-divisible} if a decomposition \eqref{Eq.DefDivisibilityPairPair} can be found in terms of intermediate CPTP maps, that is $\mathcal{N}_{k | k-1}\otimes \mathcal{I}_{d_k} (\rho)\geq 0,\, ~\forall ~\rho\geq0$ and $d_k$ is the dimension of $\mathcal{N}_{k | k-1}$'s input space. Notice that the definition of CP-divisibility is more stringent than the one for P-divisibility. CP-divisibility implies P-divisibility, so that the latter is a necessary condition for the former. Thus, we may find dynamics which are P-divisible but not CP-divisible. 

Within this context, we aim to address the essential question: How can we find whether a given channel is divisible? To illustrate,  consider the simplest case of two time steps, $t_1$ and $t_2$. In this case, the aim is to determine if a channel $\mathcal{N}_{2|0}$ can be decomposed as $\mathcal{N}_{2|0}= \mathcal{N}_{2|1} \circ \mathcal{N}_{1|0}$, given that we know $\mathcal{N}_{1|0}$. If $\mathcal{N}_{1|0}$ is invertible, there is a simple solution to this equation, namely:
\begin{equation}\label{invertible divisibility}
    \mathcal{N}_{2|1}=\mathcal{N}_{2|0} \circ \mathcal{N}_{1|0}^{-1}.
\end{equation}
In this case, we simply need to check whether $\mathcal{N}_{2|1}$ is CP, P, or neither. Although there are cases where this calculation does not bring any complication, it does not solve two problems: (i) what if $\mathcal{N}_{1|0}$ is not invertible? (ii) Given that the dynamics is not divisible, how can we quantify its degree of non-divisibility? In the next sections, we present an SDP formulation that is able to solve these questions.

\section{Conditional Quantum States and Divisibility - A Brief Review}\label{Sec.CJIsomorphism}

Here, we review the two mathematical tools we will use extensively in this work. Inspired by the approach developed in Refs.~\cite{LS13,LS14} we start this section with a brief review of the Choi-Jamio\l kowski (CJ) isomorphism. Next, we carefully define what we mean by divisibility for a given quantum dynamics---in particular, the completely positive and trace-preserving CP-divisibility and P-divisibility. We conclude this section by connecting divisibility with CJ-states, paving the road for the task of designing an efficient witness for (CP or P) divisibility. For completeness, we provide the proofs of the propositions in Appendix \ref{App.ProofIsomorphism}.

\subsection{Conditional States Approach}\label{Subsec.CondStates}

We start defining the Jamio\l koswki isomorphism~\cite{Jamiolkowski72}. In our work, it assumes the following form.

\begin{definition}[Jamio\l kowski Isomorphism]
Let a quantum system be associated with a Hilbert space $\mathcal{H}$. The set $\mathcal{L}(\mathcal{H})$ represents the linear
operators acting in $\mathcal{H}$. Let 
\begin{equation}
\mathcal{N}_{B|A}:\mathcal{L}(\mathcal{H}_{A}) \rightarrow \mathcal{L}(\mathcal{H}_{B})
\end{equation}
be a CPTP map. The \emph{(Choi-)Jamio\l kowski image} of $\mathcal{N}$ is the operator $\rho \in \mathcal{L}(\mathcal{H}_A) \otimes \mathcal{L}(\mathcal{H}_B)$ defined as
    \begin{equation}
        \rho_{B|A} := (\mbox{id}_{A} \otimes \mathcal{N}_{B|A} \circ \mbox{T}_{A}) \sum_{i,j}^{d}\ket{i}\bra{j}_{A} \otimes\ket{i}\bra{j}_{A},
        \label{Eq.DefCJIsomorphism}
    \end{equation}
where $d=\mbox{dim}(\mathcal{H}_{A})$ and the transposition $\mbox{T}_{A}$ is taken with respect to some basis in $\mathcal{H}_{A}$. On the other hand, the action of $\mathcal{N}_{B|A}$ on $\mathcal{L}(\mathcal{H}_{A})$ is given by
    \begin{equation}
        \mathcal{N}_{B|A}(\sigma_{A})=\mbox{Tr}_{A}\left[\rho_{B|A} ( \sigma_{A} \otimes \mathds{1}_{B}) \right],
    \end{equation}
where $\mathcal{N}_{B|A}(\sigma_{A}) \in \mathcal{D}(\mathcal{H}_{B})$ and the set $\mathcal{D}(\mathcal{H}_B) :=
\{\rho_{B} \in \mathcal{L}(\mathcal{H}_X)| \rho_{B} \geq 0,\text{Tr}(\rho_{B}) = 1\}$ is the state space over $\mathcal{H}_{B}$. 
\label{Def.JamilIso}
\end{definition}

\textbf{Remark:} As we mentioned before, the notation $\mathcal{N}_{B|A}$ is intended to mean a map that outputs a state on a Hilbert space $\mathcal{H}_{B}$ given that its input Hilbert space is $\mathcal{H}_{A}$. We adhere to this `$B$ given $A$' notation to create a closer parallel with refs.~\cite{LS13, LS14}. We will show how to adapt this notation to our case (with time stamps) in a minute.
 
We emphasise that contemporary literature is still debating about whether the transposition map should or should not appear in Eq.~\eqref{Eq.DefCJIsomorphism}---see Refs.~\cite{MKPM19,OCB12,CDAP09}. Different authors are more inclined towards one or another, but in this work, we will stick to the def.~\ref{Def.JamilIso} above, following Refs.~\cite{LS13, LS14}. In any case, the fact that $\rho_{B|A}$ and $\mathcal{N}_{B|A}$ are isomorphically connected can be easily checked, since
\begin{align}
    &\mbox{Tr}_{A}[\rho_{B|A}(\sigma_{A} \otimes \mathds{1}_{B})] = \nonumber \\ &=\mbox{Tr}_{A}\left[ (\mbox{id} \otimes \mathcal{N}_{B|A})(\sum_{i,j}^{d} \ketbra{i}{j} \otimes \ketbra{j}{i}) (\sigma_{A} \otimes \mathds{1}_{B}) \right] \\
    &= \sum_{i,j}^{d}\mbox{Tr}_{A}\left[ \ketbra{i}{j}\sigma_{A} \otimes \mathcal{N}_{B|A}(\ketbra{j}{i}) \right] = \sum_{i,j} \bra{j}\sigma_{A}\ket{i} \mathcal{N}(\ketbra{j}{i}) \nonumber \\
    &=\sum_{i,j}^{d} (\sigma_{A})_{j,i}\mathcal{N}(\ketbra{j}{i}) =\mathcal{N}(\sigma_{A}). \nonumber
\end{align}

The idea behind the Jamio\l kowski isomorphism is that it maps any CPTP map into a bipartite state. That is formalized in the result below, whose proof can be found in the Appendix \ref{App.ProofIsomorphism}.

\begin{proposition}
Let $\mathcal{N}_{B|A}:\mathcal{L}(\mathcal{H}_{A}) \rightarrow \mathcal{L}(\mathcal{H}_{B})$ be a linear map, and let $\rho \in \mathcal{D}(\mathcal{H}_{B} \otimes \mathcal{H}_{A})$ be the Jamio\l kowski isomorphic operator associated to it. It follows that $\rho$ satisfies
\begin{enumerate}
    \item[(a)] $\rho_{B|A}^{{T}_{A}} \geq 0$
    \item[(b)] $\mbox{Tr}_{B}[\rho_{B|A}]=\mathds{1}_{A}$
\end{enumerate}
if, and only if, $\mathcal{N}_{B|A} \circ \mbox{T}_A$ is a completely positive and trace preserving map.
\label{Prop.StateChoi}
\end{proposition}

Our starting point will generally be quantum channels rather than arbitrary linear maps. We will be interested in determining precisely when a given collection $\{\mathcal{N}_{t | 0}\}_{t \in [N]}$ of CPTP maps are divisible. In this case, Prop.~\ref{Prop.StateChoi} remains applicable, as it suffices to change item (a) from demanding that $\rho^{T_{A}}_{t|0} \geq 0$ to demanding, instead, that $\rho_{t|0} \geq 0$. So whenever working with CPTP maps, to factor in the action of the partial transposition, instead of dealing with $\rho_{t|0}$, we will consider a slightly different operator. We will use $\varrho_{t|0}:=\rho^{T_{A}}_{t|0} \geq 0$, as it results in a positive operator. In summary, $\rho_{t|0}$ is the bipartite conditional state associated with $\mathcal{N}_{t|0}$, and $\varrho_{t|0}$ is the bipartite conditional state that must be positive to render $\mathcal{N}_{t|0}$ completely positive.

Throughout this work, we will consistently use a particular index notation for states and channels alike. We will write time steps and Hilbert spaces' labels in a manner reminiscent of the ``given'' notation commonly used for conditional probabilities. Thus, when  a channel is written as $\channel{t_j}{t_i}$, then it is meant to imply, first, that  
    \begin{equation}
        \mathcal{N}_{t_j|t_i}:\mathcal{L}(\mathcal{H}_{i}) \rightarrow \mathcal{L}(\mathcal{H}_{j}),
    \end{equation}
and, secondly, that thought as an evolution map, it determines the dynamics from time step $t_i$ to time step $t_j$---with $t_j \geq t_{i}$. Analogously, the Choi-Jamio\l kowski state associated with this channel will be written as $\varrho_{t_j|t_i}$, where
    \begin{equation}
        \varrho_{t_j|t_i} \in \mathcal{L}(\mathcal{H}_{i}) \otimes \mathcal{L}(\mathcal{H}_{j}).
    \end{equation}
This notation improves the calculations and the reading of our expressions. This doubled notation also facilitates the comparison with other works where expressions like $a|b$ have a Bayesian-probabilistic meaning---see Refs.~\cite{LS13,LS14}.     

Definition \ref{Def.JamilIso} shows how to connect a single quantum channel $\mathcal{N}_{t_i|t_j}$ with its respective conditional state $\varrho_{t_i|t_j}$. It is quite natural to ask how we can calculate the composition of two channels. In other words, given $\mathcal{N}_{t_i|t_j}$ and $\mathcal{N}_{t_k|t_i}$ we may want to determine what the Jamio\l kowski image of the composition $\mathcal{N}_{t_k|t_i} \circ \mathcal{N}_{t_i|t_j}$ is. The proposition below, whose proof is in Appendix \ref{App.ProofCompositionLaw}, address this question.

\begin{proposition}
Let $\state{1}{0}$, $\state{2}{0}$, $\state{2}{1}$ be the Choi-Jamio\l kowski images of the CPTP maps $\channel{1}{0}: \mathcal{L}(\mathcal{H}_{0}) \rightarrow \mathcal{L}(\mathcal{H}_{1})$, $\channel{3}{1}: \mathcal{L}(\mathcal{H}_{0}) \rightarrow \mathcal{L}(\mathcal{H}_{2})$ and $\channel{2}{1}: \mathcal{L}(\mathcal{H}_{1}) \rightarrow \mathcal{L}(\mathcal{H}_{2})$ respectively. The composition 
    \begin{equation}
        \channel{2}{0}=\channel{2}{1} \circ \channel{1}{0}
        \label{Eq.PropMapsDiv}
    \end{equation}
holds true if and only if
    \begin{equation}
        \state{2}{0} = \mbox{Tr}_{\mathcal{H}_{1}}[(\mathds{1}_{\mathcal{H}_{0}} \otimes \state{2}{1})^{T_{\mathcal{H}_{1}}}(\state{1}{0} \otimes \mathds{1}_{\mathcal{H}_{2}})].  
        \label{Eq.StatesDiv}
    \end{equation}
\label{Prop.DivChannelDivStates}
\end{proposition}

Whenever we have to decide whether a dynamics $\mathcal{F}=\{\mathcal{N}_{k|0} \}_{k \in [N]}$ is divisible (either P or CP), Prop.~\ref{Prop.DivChannelDivStates} tell us it suffices to check out whether an equation like  Eq.~\eqref{Eq.StatesDiv} holds true for each pair of time steps $(t_i,t_{i+1})$. By doing so, instead of looking for the existence of intermediate CPTP or PTP maps verifying the composition in Eq.~\eqref{Eq.PropMapsDiv}, what can be rather difficult to establish ~\cite{BD16,RHP14,FPMZ17}, we will show that problem can be addressed by checking $N-1$ instances of an easier optimisation problem.

To establish our main result---an efficient quantifier of non-divisibility in terms of channel discrimination---we will, for simplicity, focus on only two intermediate time steps:  $t_i$ and $t_{i+1}$. The following corollary is a direct rewriting of Prop.~\ref{Prop.DivChannelDivStates}, more adequate to our framework.

\begin{corollary}
Let $\mathcal{N}_{i|0}$ and $\mathcal{N}_{i + 1|0}$ 
be two CPTP maps. 
Additionally, let $\rho_{i|0}$ and $\rho_{i+1|0}$ be respectively the 
Jamio\l kowski images of them. There is an intermediate CPTP map $\mathcal{N}_{i+1|i}$ satisfying
    \begin{equation}
        \mathcal{N}_{i+1|0}=\mathcal{N}_{i+1|i} \circ \mathcal{N}_{i|0}
        \label{Eq.WhenDynIsDiv}
    \end{equation}
if and only if the following equality holds true:
    \begin{equation}
        \varrho_{i+1|0}=\mbox{Tr}_{\mathcal{H}_{i}}\left[(\mathds{1}_{\mathcal{H}_{0}} \otimes \varrho_{i+1|i})^{T_{\mathcal{H}_{i}}} (\varrho_{i|0} \otimes \mathds{1}_{\mathcal{H}_{i+1}}) \right],
        \label{Eq.WhenDynIsDiv2}
    \end{equation}
where $\varrho_{i+1|i}$ must be the Jamio\l kowski image of $\mathcal{N}_{i+1|i}$. 
\label{Cor.WhenDynIsDiv}
\end{corollary}

\section{Divisibility as semi-definite program}\label{Sec.SDPFormulation}

It is exactly the previous corollary that allows us to reformulate the question of whether or not a given quantum dynamics $\mathcal{F}=\{\mathcal{N}_{k|0}\}_{k \in [N]}$ is P or CP divisible. Focusing on the image of the quantum channels via the Jamio\l kowski isomorphism, the question about the CP-divisibility of two given quantum maps, $\varrho_{2|0}$ and $\varrho_{1|0}$, can be re-cast as SDPs, as shown below.

\begin{tcolorbox}
    \subsection*{CP-Divisibility}
    \vspace{-1cm}
\begin{align}\label{SDP_CPTP_Divisibility}
    &\text{Given}~\varrho_{2|0}, \varrho_{1|0} \nonumber \\
    &\text{Find}~\varrho_{2|1} \nonumber \\
    &\text{s.t.}\\
    & \qquad \varrho_{2|0}=\mbox{Tr}_{B}\left[(\mathds{1}_{\mathcal{H}_{0}} \otimes \varrho_{2|1})^{T_{\mathcal{H}_{1}}} (\varrho_{1|0} \otimes \mathds{1}_{\mathcal{H}_{2}}) \right] \nonumber \\
    & \qquad \varrho_{2|1} \geq 0 \nonumber \\
    & \qquad \mbox{Tr}_{\mathcal{H}_{2}}\left[ \varrho_{2|1} \right] = \mathds{1}_{\mathcal{H}_{1}} .\nonumber
\end{align}
\end{tcolorbox}

The SDP above provides a yes/no answer to whether the map $\Lambda_{2|0}$ can be obtained as a composition of $\Lambda_{1|0}$ with an intermediate CPTP map. A variant of this problem which is also an SDP, involving the minimization of the distance between $\Lambda_{2|0}$ and $\Lambda_{2|1}\circ \Lambda_{1|0}$ for any valid CPTP $\Lambda_{2|1}$, would also provide us with an answer for how well the divisibility can be approximated. In particular, a null distance means that perfect divisibility is possible.

An operationally meaningful measure for the distance is the diamond norm \cite{Watrous2018, Watrous_NormSDP_2012}, which can be cast as an SDP and is directly related to the task of optimally distinguishing two channels when entangled resources are provided \cite{PirandolaEtAl19,BPH15} (see Appendix \ref{App.DiamondNormSDP} for details). We represent the diamond norm by $\| \bullet \|_\diamond$. An advantage of reformulating the SDP in terms of distance to divisibility is that it admits an equivalent dual form which provides, as subproducts of the optimization, a pair of operators $(W,\Lambda)$ that work together as a witness for CP-divisibility, for a fixed $\varrho_{1|0}$. They give rise to the necessary condition $\mbox{Tr}[W\,\varrho_{2|0}] + \mbox{Tr}[\Lambda] \leq 0$ for divisibility, although limited to the given pair of maps $\varrho_{2|0}$ and $\varrho_{1|0}$. Details of the dual form and witness are presented also in Appendix \ref{App.DiamondNormSDP}.

As shown below, the only modification from the SDP \eqref{SDP_CPTP_Divisibility} is the replacement of the equality requirement $\Lambda_{2|0} = \Lambda_{2|1} \circ \Lambda_{1|0}$, which can only be satisfied if $\Lambda_{2|0}$ is (P- or CP-)divisible with respect to $\Lambda_{1|0}$, with a minimization of the distance between the two sides of the equation. This replacement introduces a ``softer'' constraint that penalizes deviations from the equality constraint but keeps the problem feasible even when the maps are not exactly divisible. This formulation of the problem, stated below, is the one we employed in the results obtained in the next sections:

\begin{tcolorbox}
    \subsection*{CP-Divisibility with Diamond-Norm}
    \vspace{-1cm}
\begin{align}\label{SDP_CPTP_Div_Norm}
    &\text{Given}~\varrho_{2|0},\, \varrho_{1|0} \nonumber \\
    &\Minimise_{\varrho_{2|1}} \| \mathcal{J}^{-1}(\varrho_{2|0}) - \mathcal{J}^{-1}(\varrho_{2|1}) \circ \mathcal{J}^{-1}(\varrho_{1|0}) \|_\diamond\nonumber \\
    &\text{s.t.}\\
    & \qquad \varrho_{2|1} \geq 0 \nonumber \\
    & \qquad \mbox{Tr}_{\mathcal{H}_{2}}\left[ \varrho_{2|1}\right] = \mathds{1}_{\mathcal{H}_{1}} .\nonumber
\end{align}
where $\mathcal{J}^{-1}(\varrho_{k|0})$ is the map associated with $\varrho_{k|0}$ via the (Choi)-Jamio\l kowski isomorphism.
\end{tcolorbox}

Analogously, in order to decide about the P-divisibility of two given quantum maps $\rho_{2|0}, \rho_{1|0}$, we can reformulate the above problem, relaxing accordingly to also allow for positive (but not completely positive) linear maps.   

\begin{tcolorbox}
\subsection*{P-Divisibility}
\vspace{-1cm}
\begin{align}\label{SDP_P_Divisibility}
    &\text{Given}~\varrho_{2|0},\, \varrho_{1|0} \nonumber \\
    &\text{Find}~\varrho_{2|1} \nonumber \\
    &\text{s.t.}\\
    & \qquad \varrho_{2|0}=\mbox{Tr}_{\mathcal{H}_{1}}\left[(\mathds{1}_{\mathcal{H}_{0}} \otimes \varrho_{2|1})^{T_{\mathcal{H}_{1}}} (\varrho_{1|0} \otimes \mathds{1}_{\mathcal{H}_{2}}) \right] \nonumber \\
    & \qquad \sum_{i\neq j}\left[ \lambda_{i}(\varrho_{2|1})+\lambda_{j}(\varrho_{2|1})\right] \geq 0 \nonumber \\
    & \qquad \mbox{Tr}_{\mathcal{H}_{2}}\left[ \varrho_{2|1} \right] = \mathds{1}_{\mathcal{H}_{1}}, \nonumber
\end{align}
where $\lambda_{k}(\varrho_{2|1})$ is the $k-$th eigenvalue of the matrix $\varrho_{2|1}$
\end{tcolorbox}

A map is positive if it preserves the positive definiteness of the input states, but not necessarily preserves positivity when extended to extra degrees of freedom. The condition to ensure positivity of a map $\mathcal{P}$ is that, for any input state $|\psi\rangle$, $\mathcal{P}(|\psi\rangle\langle\psi|) \geq 0$, which can also be stated as $\langle \phi | \mathcal{P}(|\psi\rangle\langle \psi|)|\phi\rangle \geq 0$. On its Choi form, $\varrho_{\mathcal{P}}$, this condition can be equivalently put as
\begin{equation}
    \mbox{Tr}[\varrho_{\mathcal{P}}\,\sigma] \geq 0,
\end{equation} 
for all separable states $\sigma$ \cite{Bengtsson_Zyczkowski_2006}. Effectively, this means that if $\varrho$ is the image of a PTP but non-CPTP map, it must act as an entanglement witness, since some entangled state $\sigma$ will fail to preserve the positivity of the expression above. To turn the P-divisibility test into an SDP, therefore requires a description (or an approximate description) of the set of entanglement witnesses in terms of linear matrix equalities and inequalities. The connection between the generalized robustness of entanglement, an entanglement measure that can be calculated by SDP, and divisibility has also been explored in \cite{PhysRevA.96.062118}.

Although a general characterization of positive maps is expected to be a hard problem, given its connection to the problem of separability, a semidefinite program characterization of maps whose domain and range are 2-dimensional Hilbert spaces is possible, due to St{\o}rmer's theorem \cite{MM07, Stormer82, Woronowicz76}. This means replacing the positive-semidefiniteness constraint in Eq.~\eqref{SDP_CPTP_Divisibility} for $\Lambda_{2|1}$ with the decomposability relation,
\begin{equation}
\Lambda_{2|1} = \Lambda_A + \Lambda_B \circ T,
\label{eq.decomposability}
\end{equation}
where $\Lambda_{A},\,\Lambda_{B} \in \textrm{CPTP}$ and $T$ is the transposition.

With this decomposition, for quantum channels mapping qubits to qubits \footnote{Or qubits to qutrits, or qutrits to qubits for that matter} we can reformulate the SDP in~\eqref{SDP_P_Divisibility} as the following optimization program:

\begin{tcolorbox}
\subsection*{P-Divisibility (Alternative for Qubits)}
\vspace{-1cm}
\begin{align}\label{SDP_P_Divisibility_Alternative}
    &\text{Given}~\varrho_{2|0},\, \varrho_{1|0} \nonumber \\
    &\text{Find}~\varrho_{2|1},\, \sigma_{A},\, \sigma_{B} \nonumber \\
    &\text{s.t.}\\
    & \qquad \varrho_{2|0}=\mbox{Tr}_{\mathcal{H}_{1}}\left[(\mathds{1}_{\mathcal{H}_{0}} \otimes \varrho_{2|1})^{T_{\mathcal{H}_{1}}} (\varrho_{1|0} \otimes \mathds{1}_{\mathcal{H}_{2}}) \right] \nonumber \\
    & \qquad \varrho_{2|1} = \sigma_{A} + \sigma_{B} \nonumber \\
    & \qquad \sigma_{A} \geq 0  \nonumber \\
    & \qquad \sigma_{B} \mbox{ is the Jamio\l kowski image of } \Lambda_{2|1} \circ T_{1}  \nonumber \\
    & \qquad \mbox{Tr}_{\mathcal{H}_{2}}\left[ \varrho_{2|1} \right] = \mathds{1}_{\mathcal{H}_{1}}, \nonumber
\end{align}
where $\Lambda_{2|1}$ is CPTP and $T_{1}$ is the partial transposition on the first system on a given local basis.
\end{tcolorbox}

As mentioned before, using the diamond norm, we can replace the hard equation in the SDP for P-Divisibility with a smoother yet physically meaningful alternative, similar to what we did for the SDP determining the CP-Divisibility. This is the optimization problem we are using in the next examples, and it is formulated as follows:

\begin{tcolorbox}
\subsection*{P-Divisibility with Diamond Norm (Alternative for Qubits)}
   \vspace{-1cm} \begin{align}\label{SDP_P_Divisibility_AlternativeDiamondNorm}
    &\text{Given}~\varrho_{2|0},\, \varrho_{1|0} \nonumber \\
    &\Minimise_{\varrho_{2|1},\,\sigma_{A},\,\sigma_{B}} \| \mathcal{J}^{-1}(\varrho_{2|0}) - \mathcal{J}^{-1}(\varrho_{2|1}) \circ \mathcal{J}^{-1}(\varrho_{1|0}) \|_\diamond\nonumber \\
    &\text{s.t.}\\
    & \qquad \varrho_{2|1} = \sigma_{A} + \sigma_{B} \nonumber \\
    & \qquad \sigma_{A} \geq 0  \nonumber \\
    & \qquad \sigma_{B} \mbox{ is the Jamio\l kowski image of } \Lambda_{2|1} \circ T_{1}  \nonumber \\
    & \qquad \mbox{Tr}_{\mathcal{H}_{2}}\left[ \varrho_{2|1} \right] = \mathds{1}_{\mathcal{H}_{1}}, \nonumber
\end{align}
where $\Lambda_{2|1}$ is CPTP, $T_{1}$ is the partial transposition on the first system on a given local basis and $\mathcal{J}^{-1}(\varrho_{k|0})$ is the map associated with $\varrho_{k|0}$ via the (Choi)-Jamio\l kowski isomorphism.
\end{tcolorbox}

\section{Witnessing non-Markovianity in a Quantum Dynamics}\label{SDPexample}

Non-markovian effects are commonly associated with a backflow of information \cite{CKR11}. One of the most typical measures of it relies on the increase of distinguishability of two states \cite{BLP}. Differently, the formalism proposed here does not rely on any witness of the backflow of information and does not request any restriction on the dynamics. To show its relevance, we will apply it to three paradigmatic models.

The first example is a collisional model. This class of model has been initially proposed as a toy model to simulate any Markovian dynamics. Recently, by allowing the generation of correlations in the environment, collisional models have been proven to be a powerful tool in the studies of non-Markovian dynamics (for a review, see \cite{CICCARELLO20221}). In this model the environment is described by separated particles and the system interacts with each particle at once and sequentially---see Fig.~\ref{Fig.CollisionalModel}. The advantage of this model is that the three main ingredients that may lead to a non-Markovian evolution, namely the correlations in the environment, its internal dynamics, and the detailed interaction between the system and the environment can be separately fully controllable. Indeed, for qubits, collisional models can simulate any non-Markovian dynamics \cite{Rybár_2012}. 

The second example is a typical decoherence channel: the dephasing channel. Note that the analysis has been made for qubit dynamics, but the formalism is general enough to be adapted to quantum systems of any dimension.

The third example is a generalization of the second model. A random unitary evolution or a convex combination of unitary channels. This is a broad family of channels with a handy representation for higher dimensions \cite{CHRUSCINSKI20131425}.

\subsection{Collisional Model}\label{subsec.collisional}

\begin{figure}
    \centering
    \includegraphics[scale=0.25]{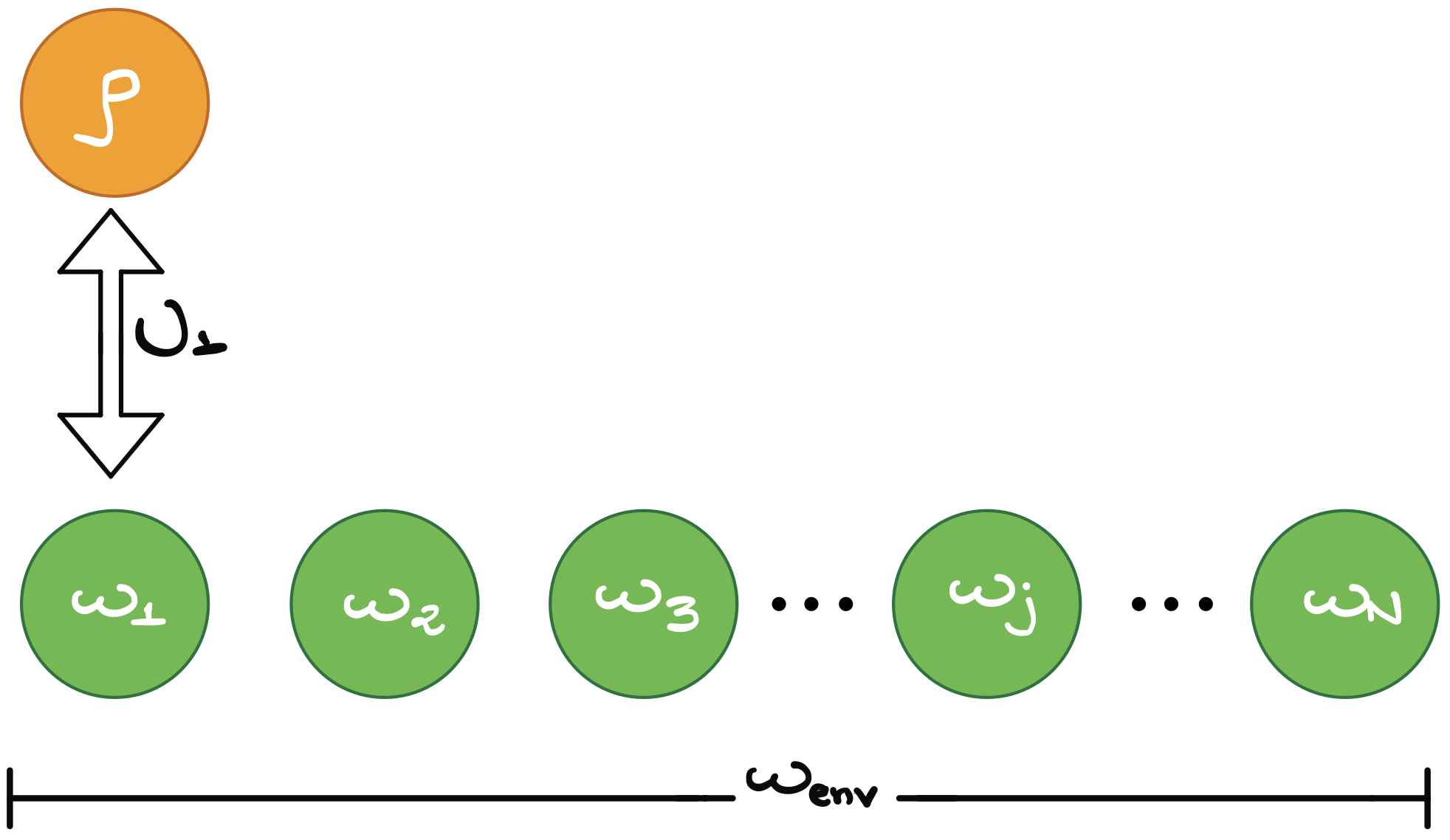}
    \caption{Collisional model's schematic representation.}
    \label{Fig.CollisionalModel}
\end{figure}

Two main ingredients are common in all collisional models:  time is discretized, and the environment is considered to be formed by elementary subsystems. Each encounter between the system and the $j$-th particle in the environment is denoted by the unitary operator $U_j$, acting for a fixed duration $\tau$. The environment's particles are typically identical and exhibit no interactions. The resulting evolution after $j$ collisions is expressed as $\rho_{j\tau|0} =\text{Tr}_j...\text{Tr}_1[U_j... U_1(\rho \otimes \omega_{env})U_1^{\dagger} ...U_j^{\dagger}]$. A pictorial representation of this process is displayed in Fig.~\ref{Fig.CollisionalModel}. This process simulates a Markovian dynamics. However, for non-Markovian dynamics, as indicated in \cite{Rybár_2012, PhysRevA.87.040103,nadjacol}, the environment's particles can no longer remain noninteracting; instead, they must exhibit some level of correlation.

Our work considers a qubit prepared in state $\rho$ that interacts with an environment from which it is initially decoupled. The interaction consists of
consecutive collisions with two particles of the environment. After the first collision, the system undergoes, with probabilities $p_X$ and $p_Z$, a transformation of $X$ or $Z$, respectively. Here $X=\ketbra{0}{1}+\ketbra{1}{0}$ and $Z=\ketbra{0}{0}-\ketbra{1}{1}$ are the Pauli matrices with $\left\{\ket{0},\ket{1} \right\}$ the computational basis. There is also a probability of $1 - p_X - p_Z$ that nothing happens to the system. After two collisions, the collisions are correlated due to correlations between the environmental particles \cite{nadjacol,BernardesEtAl15}. We focus on the case where $p_X = p_Z = p/2$, so that $1-p\,\in\,[0,1]$ is the probability that nothing happens on a single collision. For two collisions, we use the same parameter $p$, with the resulting map having almost the same functional form as that of two independent collisions. Still, we assume that correlations between collisions are such that the cross terms $X\,Z$ or $Z\,X$ are suppressed, instead returning the system to the original state.

More precisely, the maps that define the evolution are given by
\begin{align}
    \Lambda_{1|0}(\rho) &= \left(1 - p\right)\,\rho + \frac{p}{2}\,X \rho X + \frac{p}{2}\,Z \rho Z, \nonumber \\
    \Lambda_{2|0}(\rho) &= \left((1-p)^2 + p^2\right)\,\rho + p (1-p)\,\left( X \rho X + Z \rho Z\right).
    \label{Eq.CollisionalModel}
\end{align}
Recall that, in this work, we are investigating the dynamics in only two different time intervals, namely $0\rightarrow 1$ and $0\rightarrow 2$. 

\begin{figure}[!t]
    \centering
\includegraphics[width=0.9\linewidth]{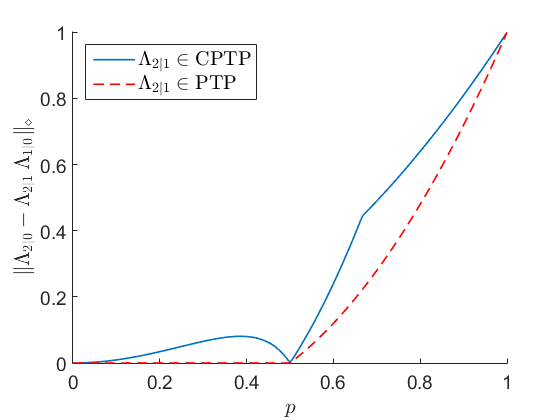}
    \caption{Deviation from exact divisibility for CPTP (solid blue curve) and PTP (dashed red curve) intermediate maps. Although the map is non-CP-divisible for almost all values of the parameter $p$, except at $p=0$, where both channels are the identity, and at $p=0.5$, where $\Lambda_{1|0}$ and $\Lambda_{2|0}$ coincide, the difference in behavior is notable between the regimes $p<0.5$ and $p>0.5$, both with the onset of a stronger form of CP-indivisibility and with P-divisibility stopping to  be attainable for $p > 0.5$.}
    \label{Fig.Comparison}
\end{figure}

Exploring the defined maps $ \Lambda_{1|0}$ and $ \Lambda_{2|0}$, we run the SDP to find the best intermediate CPTP map $ \Lambda_{2|1}$ that could approximate $ \Lambda_{2|0}$ by $\Lambda_{2|1}\Lambda_{1|0}$. In Fig.~\ref{Fig.Comparison}, we compare the distance between the total evolution and the composed evolution, $\|\Lambda_{2|0}-\Lambda_{2|1}\Lambda_{1|0}\|_\diamond$. Note that the distance is only zero if the dynamics is Markovian, which is not the case in general, except for specific values of the parameter $p$. By quantifying the distance to CP-divisibility, different regimes of non-CP-divisibility can be observed between the ranges $p<0.5$ and $p>0.5$, as a steeper curve is obtained for the latter, indicating the onset of a stronger form of non-CP-divisibility.

Furthermore, by employing the decomposability relation of Eq.~\eqref{eq.decomposability}, we also tested for P-divisibility and, more strikingly than CP-divisibility, exact P-divisibility is attainable for $p \leq 0.5$, with a clear transition to non-P-divisibility for $p>0.5$.

An advantage of using SDPs to quantify P- and CP-non-divisibility is that the best channel approximating the optimal distance is obtained as a by-product. In some cases, however, the resulting best channel has been observed to be the identity, meaning that no other channel will improve in approximating $\Lambda_{2|0}$ starting at $\Lambda_{1|0}$, $\Lambda_{1|0}$ being already the closest channel as a factor composing the former. This induces a notion of absolute non-divisibility and a possible justification for the result is introduced in Appendix \ref{App.AbsoluteNondiv}.

\begin{figure}[!t]
    \centering
    \includegraphics[width = \linewidth]{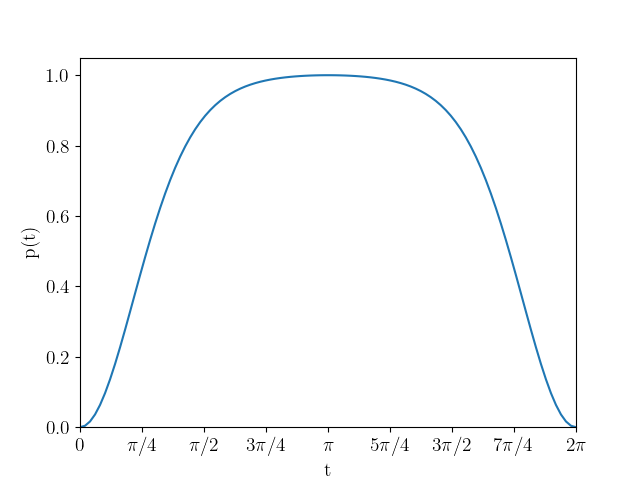}
    \caption{Dephasing parameter as a function of time.}
    \label{fig:Dephasing_prob}
\end{figure}

\begin{figure*}[!t]
    \centering
    \includegraphics[width = \columnwidth]{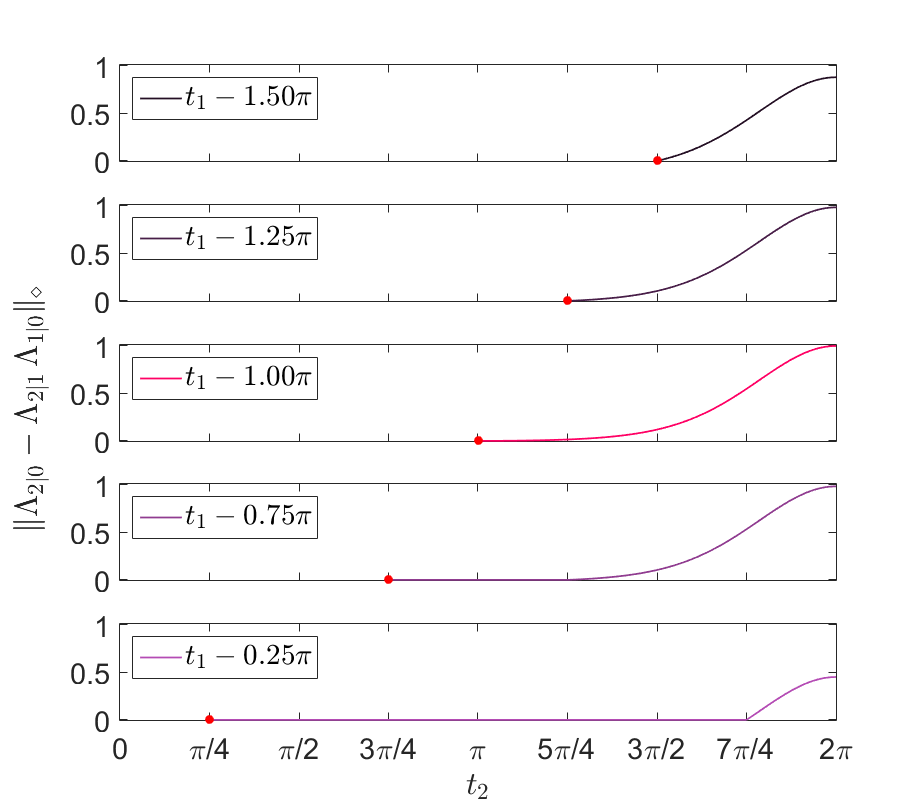}
    \includegraphics[width = \columnwidth]{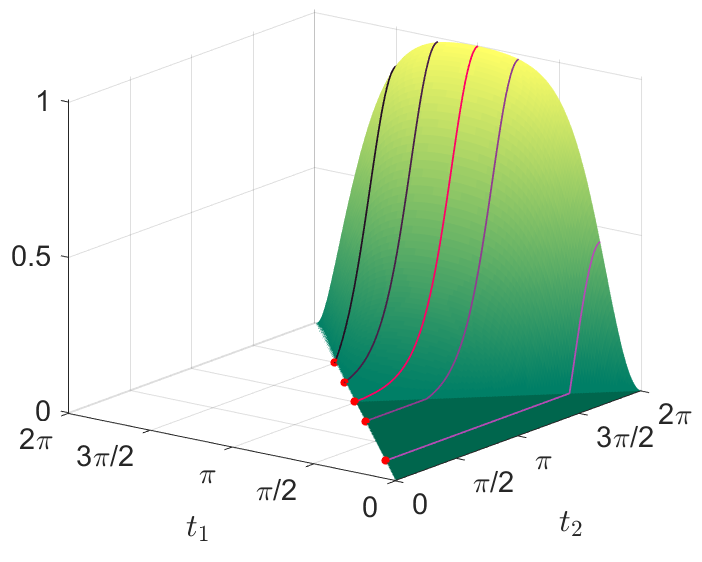}
    \caption{Distance to divisibility for the dephasing channel. Dynamics starts at time $t_1$ and ends at time $t_2$ (with values in the range $[0, 2\pi]$ and $t_2 \geq t_1$). On the left panel, divisibility for a selection of initial times is shown. Notice that, as $t_1$ increases, the sooner non-divisibility starts, with extended regions of exact divisibility occurring only for $t_1 \leq \pi$. Compared with figure \ref{fig:Dephasing_prob}, it can be seen that divisibility can only be attained when $p(t_2) > p(t_1)$.}
    \label{fig:Dephasing}
\end{figure*}

\subsection{Dephasing Dynamics}\label{subsec.dephasing}

The dephasing channel is a completely quantum noise channel, where quantum information is lost without the loss of energy. It models fiber-optic communication channels \cite{Erhard20} and superconducting circuits \cite{Brito_2008}, impacting quantum communication and quantum computation.

We start considering the divisibility for the dephasing map, given by 
\begin{equation}
\Lambda_t(\rho) = \left(1 - \frac{p(t)}{2}\right)\rho + \frac{p(t)}{2} Z\,\rho\,Z,    
\label{eq:dephasing}
\end{equation}
with $p(t) = e^4\,(1 - e^{-2(1-\cos(t))})/(e^4 - 1)$ (see Fig.~\ref{fig:Dephasing_prob}). The corresponding master equation is $\dot{\rho}=\frac{\gamma(t)}{2}(Z\rho Z-\rho)$ with $\gamma(t)\leq 0$ for $t \geq \pi$ and consequently non-Markovian \cite{CHRUSCINSKI20131425}. Another interesting feature of this model is that for $p=1$ the map is non-invertible; this is a regime where many different Markovian witnesses are not well defined \cite{Dario1,Dario2,Janek}. The time parameter here is again continuous, but note that we will always be interested in three different time steps: the initial time $t_0$, the intermediate time $t_1$, and the final time $t_2$.

Using the SDP test described in Eq.\ \eqref{SDP_CPTP_Div_Norm}, we analysed the possibility of obtaining a CPTP intermediate map $\Lambda_{2|1}$ that could connect the maps at times $t_1$ and $t_2$ (such that $\Lambda_{2|0} = \Lambda_{2|1}\,\circ\,\Lambda_{1|0}$), where $\Lambda_{2|0}=\Lambda_{t_2}$ and $\Lambda_{1|0}=\Lambda_{t_1}$ in Eq.~\eqref{eq:dephasing}.

The results are presented in Fig. \ref{fig:Dephasing}, each point $(t_2, t_1)$ corresponding to the minimal distance $\|\Lambda_{2|0} - \Lambda_{2|1} \circ \Lambda_{1|0} \|_{\diamond}$ overall CPTP maps $\Lambda_{2|1}$, meaning that exact divisibility is only attained when the distance is zero. In all other points, the final map $\Lambda_{2|0}$ can only be approximated from the initial map $\Lambda_{1|0}$, with a better quality of approximation for decreasing values of the minimal distance. This can be understood if we observe Fig.~\ref{fig:Dephasing_prob}. It is clear that if $p(t_2) \geq p(t_1)$ for different instants of time $t_2$ and $t_1$, then $\Lambda_{2|0}$ can be divided by $\Lambda_{1|0}$, by compositing it with another dephasing map. In the opposite case, for $p(t_2) < p(t_1)$, Fig.\ \ref{fig:Dephasing} shows that exact divisibility is no longer attainable and an approximation $\Lambda_{2|0} \approx \Lambda \circ \Lambda_{1|0}$ is the best that one can hope to obtain. This approximation worsens for larger differences of $p(t_2) - p(t_1)$.

In systems with dimension higher than two, one possible way to generalize the dephasing channel is given by replacing the random action of the Pauli $Z$ unitary with projections on the elements of the computational basis, $\Pi_k = \ketbra{k}{k}$ \cite{Lautenbacher_2024}. The effect of the dephasing channel on a density matrix $\rho$ is given by
\begin{equation}
\Lambda^{\textrm{dep}}_p(\rho) = (1 - p)\rho + p\,\sum_k \Pi_k\,\rho\,\Pi_k,
    \label{eq:dephasing_hd}
\end{equation}
where the channel is now directly parametrized by the probability of getting a projection, $p \in [0,\,1]$.

\begin{figure*}[!t]
    \centering
    \includegraphics[width = \linewidth]{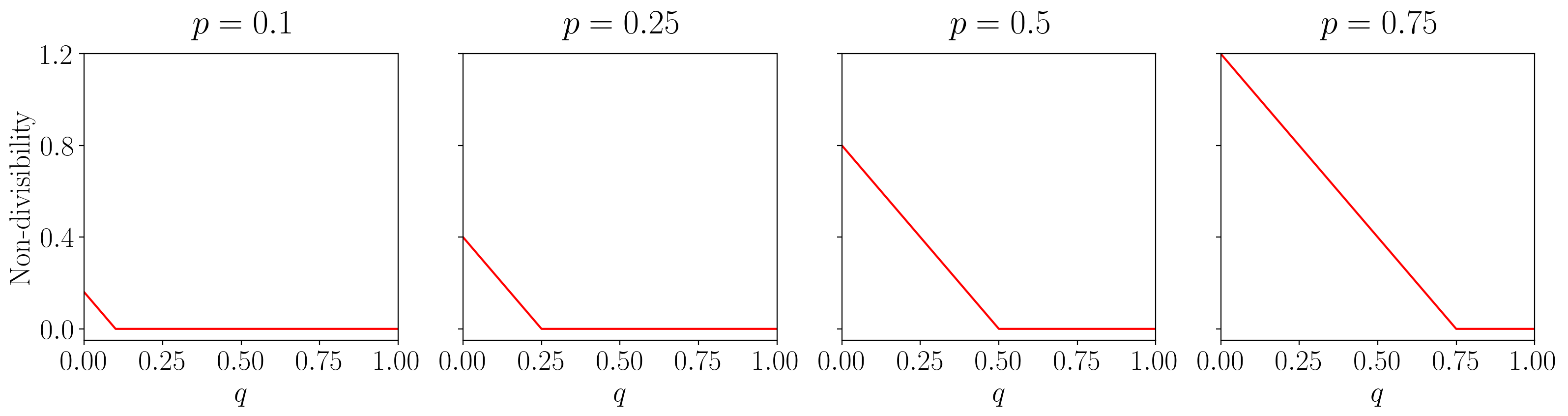}
    \caption{CP-nondivisibility for different values of parameters $p$ and $q$ in a space with dimension $5$. The onset of divisibility occurs exactly at $q = p$, with the resulting intermediate map confirmed numerically to be $\Lambda_{2|1} = \Lambda^{\textrm{dep}}_{(q-p)/(1-p)}$.}
    \label{fig:Dephasing_nondiv_hd}
\end{figure*}

Notably, also here the composition of two dephasing channels return a third dephasing channel, with dephasing probability greater than the maximum of the two input probabilities. Direct calculation shows that $\Lambda_{p'} \circ \Lambda_{p} = \Lambda_{p + p' - p p'}$, and $p+p'-p p' \geq \max\{p, p'\}$. This implies in particular that, for $\Lambda_{1|0} = \Lambda^{\textrm{dep}}_p$ and $\Lambda_{2|0} = \Lambda^{\textrm{dep}}_q$, a solution to the divisibility problem is always obtainable when $q \geq p$, by using $p' = (q - p)/(1-p)$, for $p < 1$, and setting $p'=0$, for $p=1$. Indeed, in running the program of Eq.\ \eqref{SDP_CPTP_Div_Norm} for channels $\Lambda_{1|0} = \Lambda^{\textrm{dep}}_p$ and $\Lambda_{2|0} = \Lambda^{\textrm{dep}}_q$ in input dimensions $3$ and $5$, the maps obtained as a result of the optimization coincide with the analytical expectation. In the cases where CP-divisibility is not achieved, namely for $q < p$, the identity channel is obtained, indicating a strong character of non-invertibility from channels with higher values for the dephasing parameter to lower values. Results for CP-nondivisibility for dimension $5$ are shown in Fig. \ref{fig:Dephasing_nondiv_hd}, where it can be observed both that $\| \Lambda_{2|0} - \Lambda_{2|1} \circ \Lambda_{1|0} \|_\diamond = \| \Lambda_{2|0} - \mathcal{I}\circ\Lambda_{1|0}\|_\diamond = (q-p)\,\| \mathcal{I} - \sum_k \Pi_k (\cdot) \Pi_k\|_\diamond$ for $q < p$, and equal to zero, within numerical precision, for $q \geq p$.

\subsection{Convex combinations of unitary channels}

Moving away from physically motivated channels, in this section we consider a class of channels that is more interesting from a mathematical perspective to provide an analysis of scalability of the method. Namely, we consider the particular subset of unital maps formed as convex combinations of unitary channels. It is a relevant and broad family of channels that encompasses also physically relevant maps, the dephasing channels of the previous subsection being a notable example thereof \cite{Watrous2018}.

In full generality, the maps are defined as
\begin{equation}
    \Lambda_{\boldsymbol{p},\,\boldsymbol{U}}(\rho) = \sum_{i=1}^n p_i\,U_i\,\rho\,U^\dagger_i,
\end{equation}
for a set of unitaries $\boldsymbol{U} = \{U_1,\ldots,U_n\}$ and a given probability distribution $\boldsymbol{p}$. In the particular cases considered here, we analyse the situation where $\Lambda_{2|0} = \mathcal{I}$ and $\Lambda_{1|0}$ is a randomly generated channel with unbiased combinations of the unitaries: $\boldsymbol{p} = \{1/n,\ldots,1/n\}$, and $U_i$ are drawn using the Haar measure.

\begin{figure}[!t]
    \centering
    \includegraphics[width = 0.9\linewidth]{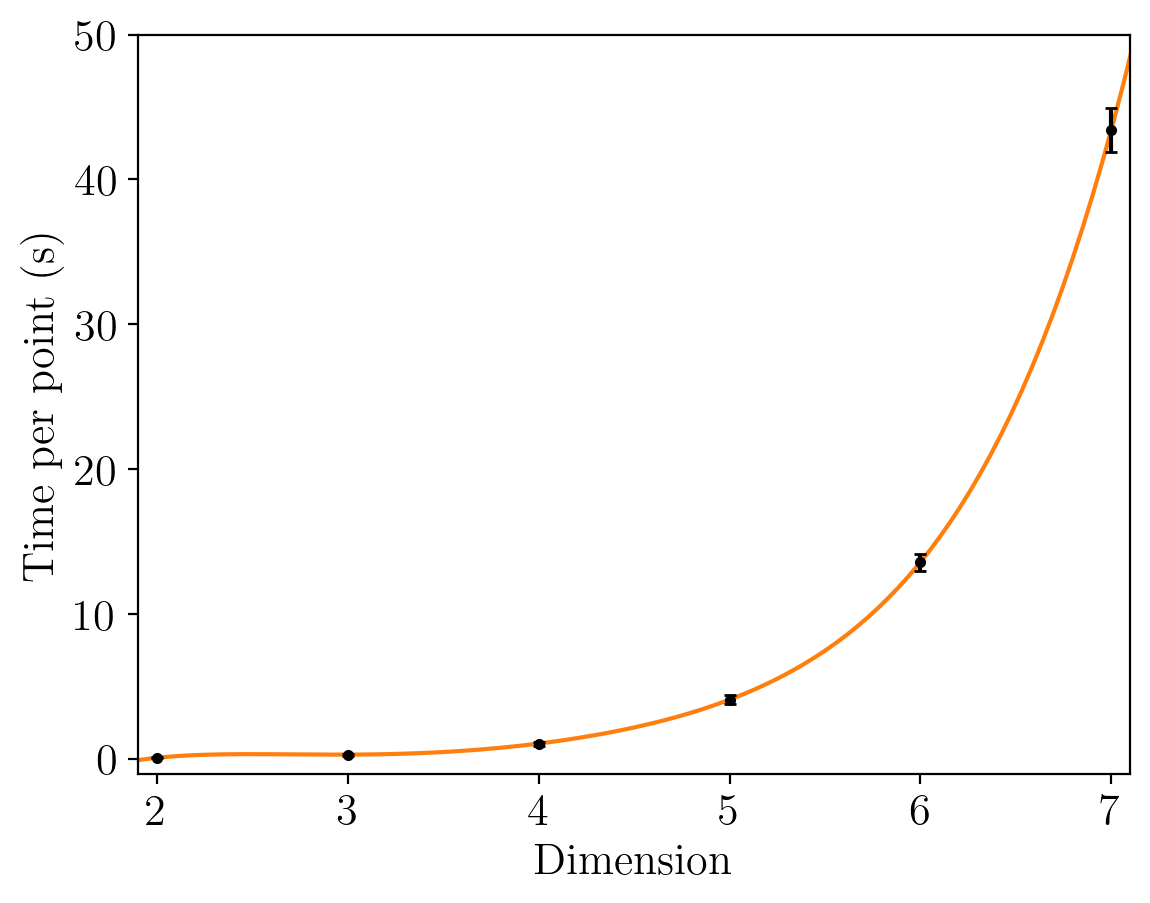}
    \caption{Average time taken to run the SDP for CP-nondivisibility for each channel $\Lambda_{1|0}$ sampled. A behavior scaling with $d^5$ is observed in practice, as shown by the fitted polynomial over the computed data.}
    \label{fig:unitaries_polyfit}
\end{figure}

The SDP is run in scenarios with differing values of the parameters. We considered Hilbert spaces of dimensions $d=2,\ldots,7$ and, for dimensions $2$ and $3$, we analyse the impact of using different number of unitaries that conform the channel, $n$. For each particular selection of parameters, $500$ channels $\Lambda_{1|0}$ are sampled and the corresponding CP-nondivisibility is computed. The optimization for each channel is timed, average values obtained for $n=5$ and for different values of the dimension are shown in Fig. \ref{fig:unitaries_polyfit}. The values are compatible with a polynomial scaling, with a practical scaling of  order $d^5$, as observed in the agreement with the curve fitted over the computed data, shown in the figure.

With respect to changes in $n$, no noticeable change is observed, as shown in table \ref{tab:times_diff_n}, which reveals that, for typical cases, particular instances of the channels used do not impact the computation of the SDP. All computations were performed on an Intel Xeon W-2295 workstation operating at $3.0$ GHz, using the MOSEK optimizer \cite{Mosek}.\\

\begin{table}[h]
    \centering
    \begin{tabular}{|c|c|c|c|c|}
        \hline
        \backslashbox{$d$}{$n$} & $1$ & $2$ & $5$ & $15$ \\
        \hline
        $2$ \rule[-2ex]{0pt}{5ex} & $0.11 \pm 0.03$ & $0.10 \pm 0.02$ & $0.10 \pm 0.03$ & $0.10 \pm 0.03$ \\
        \hline
        $3$ \rule[-2ex]{0pt}{5ex} & $0.30 \pm 0.06$ & $0.33 \pm 0.07$ & $0.29 \pm 0.06$ & $0.32 \pm 0.06$ \\
        \hline
    \end{tabular}
    \caption{Average time taken to solve each SDP in seconds, for different values of $n$ and dimensions $d=2,\,3$. Values are comparable for fixed dimensions, showing that the technique is not affected by particular instances of input channels used.}
    \label{tab:times_diff_n}
\end{table}

\section{Conclusion}\label{Sec.Conclusion}

In this work, we have designed a method to decide whether any given quantum dynamics fails to be CP-divisible or P-divisible. The essential feature of our formalism is that instead of addressing the divisibility problem directly from the perspective of quantum channels, we first switch pictures and approach the situation from the standpoint of composability of quantum states, which can be done via the Choi-Jamio\l kowski isomorphism. It is precisely this apparently naive switching of perspective that grants power to our approach. 

Unlike other methods found in the literature, our toolbox can be cast as a single SDP that decides whether a given family of CPTP maps is CP-divisible and returns the passage maps (if any) connecting any two consecutive instants of time. Additionally, by weakening the formulation of our program, we can similarly investigate the P-divisibility of any given discrete quantum dynamics. In the cases where the dynamics is not CP-divisible, upon a small modification, we can also adapt our method and seek for the best (in the diamond-norm) CPTP map that connects two consecutive instants of time.  Finally, we emphasize that our method is applied to any given family of CPTP maps, and artificial requirements of dimensionality or invertibility do not plague it. As an illustration of our method, we analyzed two paradigmatic quantum dynamics in the literature, the dephasing and collisional models. In the collisional case, our quantifier shows a clear difference between the CPTP and P (non-)divisibility. This enables us to identify two different non-Markovian regimes, called weak (P-divisible) and strong non-Markovian (non-P-divisible) \cite{sabrina}. In the dephasing dynamics, we were able to identify non-divisibility for the case where the map does not have an inverse. Moreover, we could identify large regions of CPTP-divisibility and also pinpoint the dynamics parameters leading to stronger non-divisibility.

We draw attention to Refs. \cite{StormerEtAl18} and \cite{AB19}. In the first contribution, the authors center attention on those cases where the generators giving rise to dynamical maps are not invertible. They are able to provide necessary and sufficient conditions in which a given quantum dynamics is divisible---their definition of divisibility embraces completely positive, positive, and functional linear divisibility. In the second contribution, the authors define a measure of non-Markovianity based on robustness. Although related to our work, there are some differences worth pointing out. In Ref.~\cite{StormerEtAl18}, results are tailored to continuous quantum dynamics, and the relation between what they obtain and those results involving discrete evolutions is not evident and deserves further analysis---a fact that is emphasized by the authors themselves. In Ref.~\cite{AB19}, the situation is similar. In order to keep their problem in the convex regime, the authors had to focus on a restricted subset of all Choi matrices; those in which the consecutive instants of time are sufficiently close together. In comparison, our approach works for discrete families of CPTP maps, but our method is efficient and versatile and can be adapted to numerically investigate continuous evolutions, provided that we discretize the dynamics.

To conclude, we discuss possible venues of investigation hinted by our results. In Ref.~\cite{Lautenbacher2022}, the authors have also considered dephasing dynamics but in the context of finding the optimal recovery map. We conjecture that instead of looking for the exact recovery map, one could get inspired by our ideas and seek the map that would best approximate the reverse map according to some meaningful norm. In this new perspective, the problem would amount to solving the SDP with the final map "$\Lambda_{2|0}$" given by the identity, which corresponds to minimizing the distance between $\Lambda' \circ \Lambda$ and the identity map over all CPTP maps $\Lambda'$, given an input CPTP map $\Lambda$. If $\Lambda$ is reversible, then $\Lambda' = \Lambda^{-1}$; otherwise, we would find its best approximation according to some norm. That is, one could not only determine whether such a recovery map exists but also exhibit its best approximation. In turn, by considering the dual version of the associated SDP, it should be possible to construct explicit witnesses for non-divisibility, that could be particularly relevant to be applied in experimental setups. These are two interesting research venues opened by this work and we hope our results might trigger further research in these directions.

\begin{acknowledgments}
Cristhiano Duarte thanks for the hospitality of the Institute for Quantum Studies at Chapman University. CD has been funded by an EPSRC grant. This research was partially supported by the National Research, Development and Innovation Office of Hungary (NKFIH) through the Quantum Information National Laboratory of Hungary and through the grant FK 135220. This research was also supported by the Fetzer Franklin Fund of the John E.\ Fetzer Memorial Trust and by grant number FQXi-RFP-IPW-1905 from the Foundational Questions Institute and Fetzer Franklin Fund, a donor-advised fund of Silicon Valley Community Foundation. RC acknowledges the Serrapilheira Institute (Grant No. Serra-1708-15763), the Simons Foundation (Grant Number 1023171, RC) and the Brazilian National Council for Scientific and Technological Development (CNPq, Grant No.307295/2020-6). R.N. acknowledges support from the Government of Spain (Severo Ochoa CEX2019-000910-S and TRANQI), Fundació Cellex, Fundació Mir-Puig, Generalitat de Catalunya (CERCA program) and support from the Quantera project Veriqtas. N.K.B. acknowledges financial support from CNPq Brazil (Universal Grant No. 406499/2021-7) and FAPESP (Grant 2021/06035-0). N.K.B. is part of the Brazilian National Institute for Quantum Information (INCT Grant 465469/2014-0).
 
\end{acknowledgments}

\appendix

\section{Proof of Proposition\ref{Prop.StateChoi}}\label{App.ProofIsomorphism}

\begin{proof}
The ``if'' part. Suppose that $\rho$ verifies (a) and (b) above. We must prove that its isomorphic image $\mathcal{N} \circ \mbox{T}_{A}$ is completely positive and trace-preserving. 
Firstly, given $\sigma_{A} \in \mathcal{D}(\mathcal{H}_{A})$ note that:
    \begin{align}
        \mbox{Tr}_{B}[\mathcal{N}_{B|A}(\sigma_{A}^{T})] &= \mbox{Tr}_{B}\mbox{Tr}_{A}[\rho_{B|A}(\sigma_{A}^{T} \otimes \mathds{1}_{B}  )] \nonumber \\
        &= \mbox{Tr}_{A}\mbox{Tr}_{B}[\rho_{B|A}( \sigma_{A}^{T} \otimes \mathds{1}_{B})] \nonumber \\
        &= \mbox{Tr}_{A}(\sigma_{A}^{T})= \mbox{Tr}_{A}(\sigma_{A}).
    \end{align}
So that $\mathcal{N}_{B|A} \circ \mbox{T}_{A}$ is trace-preserving. Now, it remains to prove that $\mathcal{N}_{B|A}\circ \mbox{T}_{A}$ is completely positive. Take $\sigma_{A} \in \mathcal{D}(\mathcal{H}_{A})$, then:
    \begin{align}
    (\mathcal{N_{B|A}}&\circ \mbox{T}_{A})(\sigma_{A}) = \nonumber \\
         &= \mbox{Tr}_{A}\left[\rho_{B|A}( \sigma_{A}^{T} \otimes \mathds{1}_{B})\right] = \mbox{Tr}_{A}\left[\sqrt{\rho_{B|A}^{T_{A}}}( \sigma_{A} \otimes \mathds{1}_{B})\sqrt{\rho_{B|A}^{T_{A}}}\right] \nonumber \\
        &=\sum_{a}(\bra{a} \otimes \mathds{1}_{B})\left(\sqrt{\rho_{B|A}^{T_{A}}}(\sigma_{A} \otimes \mathds{1}_{B})\sqrt{\rho_{B|A}^{T_{A}}}\right) (\ket{a} \otimes \mathds{1}_{B}) \nonumber \\
        &=\sum_{a}\left[(\bra{a} \otimes \mathds{1}_{B})\sqrt{\rho_{B|A}^{T_{A}}}\right](\sigma_{A} \otimes \mathds{1}_{B})\left[\sqrt{\rho_{B|A}^{T_{A}}}(\ket{a} \otimes \mathds{1}_{B})\right]
    \end{align}
Defining, $K_{a}:=[(\bra{a} \otimes \mathds{1}_{B})\sqrt{\rho_{B|A}^{T_{A}}}]$ for all $a$, we can rewrite $\mathcal{N}_{B|A}(\sigma_{A})$ as
\begin{equation}
    \mathcal{N}_{B|A}(\sigma_{A})=\sum_{a}K_{a}(\sigma_{A} \otimes \mathds{1}_{B})K_{a}^{\ast},
\end{equation}
and that concludes the first half of the proof. 

The ``only if'' part. Assume that $\mathcal{N}_{B|A} \circ \mbox{T}_{A}$ is completely positive and trace-preserving. For this part, we use the inversion expressed in eq.~\eqref{Eq.DefCJIsomorphism}. As a matter of fact,
\begin{align}
    \mbox{Tr}_{B}(\rho_{B|A}) &= \mbox{Tr}_{B}\left[\sum_{i,j}^{d} \ket{i}\bra{j} \otimes \mathcal{N}_{B|A}(\ketbra{j}{i})\right]  \nonumber \\
    &= \sum_{i,j}^{d} \ket{i}\bra{j} \otimes \mbox{Tr}\left[\mathcal{N}_{B|A}\circ \mbox{T}_{A}(\ketbra{i}{j})\right]  \\
    &= \sum_{i,j}^{d} \ket{i}\bra{j}[\mbox{Tr}\ketbra{j}{i})] \nonumber \\
    &=\sum_{i,j}^{d}\ketbra{i}{j}\delta_{i,j}=\mathds{1}_{A}, \nonumber
\end{align}
and that is saying that Tr$_{B}(\rho_{B|A})=\mathds{1}_{A}$. Now, we have to prove item (a). That is done by noticing that $\rho_{B|A}^{T_{A}} = \mbox{id} \otimes (\mathcal{N}_{B|A}\circ \mbox{T}_{A}
)(\Phi^{+})$, where $\Phi^{+}:=\sum_{i,j}\ketbra{i}{j} \otimes \ketbra{i}{j}$ is the usual (non-normalised) Bell-state. As the composition $\mathcal{N}_{B|A} \circ \mbox{T}_{A}$ is completely positive, it implies that $\rho_{B|A} \geq 0$.
\end{proof}

\section{Proof of Proposition \ref{Prop.DivChannelDivStates}}\label{App.ProofCompositionLaw}

\begin{proof}
The "if" part. Assume that eq.~\eqref{Eq.PropMapsDiv} holds true. In this case:
\begin{equation}
    \channel{2}{0}=\channel{2}{1} \circ \channel{1}{0}: \mathcal{L}(\mathcal{H}_{0}) \rightarrow \mathcal{L}(\mathcal{H}_{2})
\end{equation}
is a completely positive trace-preserving map arising from the composition of the other two CPTP maps. For the sake of comprehension, denote $\rho_{i|j}$ as $\mathcal{J}(\mathcal{N}_{i|j})$. Then,
\begin{align}
    \varrho_{2|0} \circ \mbox{T}_{\mathcal{H}_{0}} &=  \rho_{2|0} = \mathcal{J}(\mathcal{N}_{2|0}) =   \mathcal{J}(\mathcal{N}_{2|1} \circ \mathcal{N}_{1|0}) \nonumber \\
    &=\left[\mbox{id} \otimes (\mathcal{N}_{2|1} \circ \mathcal{N}_{1|0})\right]\left(\sum_{i,j}\ketbra{i}{j} \otimes \ketbra{j}{i}\right) \nonumber \\
    &= (\mbox{id} \otimes \mathcal{N}_{2|1})(\mbox{id} \otimes \mathcal{N}_{1|0})\left(\sum_{i,j}\ketbra{i}{j} \otimes \ketbra{j}{i}\right)  \\
    &= (\mbox{id} \otimes \mathcal{N}_{2|1})(\rho_{1|0}) \nonumber \\
    & = \mbox{Tr}_{\mathcal{H}_{1}}\left[ (\mathds{1}_{\mathcal{H}_{0}} \otimes \rho_{2|1}) ( \rho_{1|0} \otimes \mathds{1}_{\mathcal{H}_{2}}) \right], \nonumber \\
    & = \mbox{Tr}_{\mathcal{H}_{1}} \left[ (\mathds{1}_{\mathcal{H}_{0}} \otimes \rho_{2|1}) ( \varrho_{1|0} \otimes \mathds{1}_{\mathcal{H}_{2}})^{T_{\mathcal{H}_{0}}} \right] \nonumber \\
    & = \left\lbrace\mbox{Tr}_{\mathcal{H}_{1}} \left[ (\mathds{1}_{\mathcal{H}_{0}} \otimes \rho_{2|1}) ( \varrho_{1|0} \otimes \mathds{1}_{\mathcal{H}_{2}}) \right]\right\rbrace^{T_{\mathcal{H}_{0}}}.
\end{align}
In conclusion, %
\begin{equation}
    \varrho_{2|0}=\mbox{Tr}_{\mathcal{H}_{1}} \left[ (\mathds{1}_{\mathcal{H}_{0}} \otimes \varrho_{2|1})^{T_{\mathcal{H}_{1}}} ( \varrho_{1|0} \otimes \mathds{1}_{\mathcal{H}_{2}}) \right].
\end{equation}

The "only if" direction follows in complete analogy, and we will not write it down here. 

\end{proof}

\section{Operational significance and SDP characterization of the diamond norm}
\label{App.DiamondNormSDP}

The diamond norm, also known in the literature as the completely bounded trace norm, is a measure of the ability to distinguish between two channels when generic states, possibly entangled with auxiliary degrees of freedom, are provided. It is defined over the space of linear superoperators, i.e. maps of the form $\Phi: \mathcal{L}(\mathcal{H}_A) \rightarrow \mathcal{L}(\mathcal{H}_B)$ that act linearly on the operators in $\mathcal{L}(\mathcal{H}_A)$, and can be expressed as
\begin{equation}
\|\Phi\|_\diamond = \max_{\substack{X \in \mathcal{L}(\mathcal{H}_d \otimes \mathcal{H}_A)\\ \|X\|_1 \leq 1}} \|I_d \otimes \Phi (X) \|_1,
\label{app_eq:diamond_norm_def}
\end{equation}
where $\mathcal{H}_d$ is a $d$-dimensional Hilbert space, with $d = \mathrm{dim}(\mathcal{H}_A)$, $I_d$ is the identity map on $\mathcal{H}_d$, and $\|\bullet\|_1$ is the trace norm for linear operators, equal to the sum of the singular values of the operator. In fact, it can be shown that any extension of $\mathcal{H}_A$ to $\mathcal{H}_{d'} \otimes \mathcal{H}_A$ with dimension $d' > d$ cannot perform better than the value for $\|\Phi\|_\diamond$ in the maximisation above \cite{Watrous2018}. Consequently, it is indifferent to include a maximisation over $d'$ on the definition above. It can also be shown that the optimum above can be attained with a rank-1 $X$ of the form $|u\rangle\langle v|$ for some normalised vectors $|u\rangle,\,|v\rangle \in \mathcal{H}_d\otimes\mathcal{H}_A$. For Hermiticity-preserving maps $\Phi$, $|v\rangle = |u\rangle$.

With the definition above, it is clear that the diamond norm is relevant for the task of distinguishing two arbitrary channels $\Psi_0$ and $\Psi_1$: Assume that an experimenter is able to prepare any state $\rho \in \mathcal{D}(\mathcal{H}_{d'} \otimes \mathcal{H}_A)$, of which they can use the subsystem on $\mathcal{H}_A$ to go through a under-characterised channel $\Psi$. It is only known that the channel $\Psi$ may be $\Psi_0$ with probability $p$, or $\Psi_1$ with probability $1-p$. Both $\Psi_0$ and $\Psi_1$ are known, as well as the probability $p$. Only which channel is actually in effect at a time is unknown. Let $\Psi'_0 \coloneqq I_{d'} \otimes \Psi_0$ and $\Psi'_1 \coloneqq I_{d'} \otimes \Psi_1$, for the input state $\rho$, the task reduces to measuring the output state and distinguishing between $\rho_0 \coloneqq \Psi'_0(\rho)$ and $\rho_1 \coloneqq \Psi'_1(\rho)$ with maximum probability.

From the Holevo-Helstrom theorem, it is known that the best measurement for the task results in a probability given by
\begin{align}
    P^{\textrm{opt}}_{\textrm{Guess}} &\coloneqq \max_{\{M_0,\, M_1\}} p\,\mathrm{Tr}[M_0\,\rho_0] +  (1-p)\,\mathrm{Tr}[M_1\,\rho_1]\nonumber \\
    &= \frac{1}{2} + \frac{1}{2}\| p\,\Psi'_0(\rho) - (1-p)\,\Psi'_1(\rho)\|_1.
\end{align}
Using definition \eqref{app_eq:diamond_norm_def}, a subsequent optimisation over the state $\rho$ results in the diamond norm of the superoperator $p\,\Psi'_0 - (1-p)\,\Psi'_1$ on the right-hand side of the equation above, which reveals that the norm measures the success of the task when the best strategy both for preparation of the state and for measurement are used.

The diamond norm can be alternatively characterised as 
\begin{equation}
\|\Phi\|_\diamond = \max_{\rho_0,\, \rho_1 \in \mathcal{D}(\mathcal{H}_A)} \| \left( \sqrt{\rho_0} \otimes \mathbb{1}_{\mathcal{H}_B} \right)
\,\mathcal{C}_\Phi\,\left(\sqrt{\rho_1} \otimes \mathbb{1}_{\mathcal{H}_B}\right)\|_1,
\end{equation}
where $\mathcal{C}_\Phi \in \mathcal{L}(\mathcal{H}_A \otimes \mathcal{H}_B)$ is the Choi map corresponding to $\Phi$ [Eq.\ \eqref{Eq.DefCJIsomorphism}] (without the transposition). The characterisation above admits a formulation in terms of semidefinite programming as follows:
\begin{align} 
    \Minimise&\quad \frac{1}{2}(u_0 + u_1) \nonumber \\
    \textrm{s.t.}&\quad u_0\,\mathbb{1} \geq \mathrm{Tr}_{\mathcal{H}_B}[Y_0] \nonumber \\
    &\quad u_1\,\mathbb{1} \geq \mathrm{Tr}_{\mathcal{H}_B}[Y_1] \nonumber \\
    &\quad \begin{bmatrix}
    Y_0 & -\mathcal{C}_\Phi \cr
    -\mathcal{C}^\dag_\Phi & Y_1
    \end{bmatrix} \geq 0, 
    \label{eq:app_diamond_norm_sdp}
\end{align}
for $Y_0,\,Y_1 \in \mathcal{L}(\mathcal{H}_A \otimes \mathcal{H}_B)$. The scalar variables $u_0$ and $u_1$ can be understood as $\|\mathrm{Tr}_{\mathcal{H}_B}[Y_0]\|_\infty$ and $\|\mathrm{Tr}_{\mathcal{H}_B}[Y_1]\|_\infty$, respectively. There is no need to introduce inequalities $\|\mathrm{Tr}_{\mathcal{H}_B}[Y_x]\|_\infty \geq -u_x\,\mathbb{1},\,(x=0,1)$, since the last matrix inequality constraint implies also $Y_0,\,Y_1 \geq 0$, and thus that their eigenvalues are nonnegative.

More explicitly, one possible standard form for a semidefinite program can be characterized by a triple $(\Xi, C, D)$ \cite{Watrous2018, Watrous_NormSDP_2012}, such that $\Xi$ is a superoperator mapping operators in $\mathcal{L}(\mathcal{V})$ to operators in $\mathcal{L}(\mathcal{W})$, and $C \in \mathcal{L}(\mathcal{V})$, $D \in \mathcal{L}(\mathcal{W})$ are Hermitian operators acting on the respective vector spaces $\mathcal{V}$ and $\mathcal{W}$. Its primal formulation is given by
\begin{align} \label{eq:app_std_form_primal}
    \Maximise_{X\,\in\,\mathcal{L}(\mathcal{V})}&\quad \mathrm{Tr}[C\,X] \nonumber \\
    \textrm{s.t.}&\quad \Xi(X) = D, \nonumber \\
    &\quad X \geq 0,
\end{align}
with a respective dual formulation given by
\begin{align} \label{eq:app_std_form_dual}
    \Minimise_{Y\,\in\,\mathcal{L}(\mathcal{W})}&\quad \mathrm{Tr}[D\,Y] \nonumber \\
    \textrm{s.t.}&\quad \Xi^\dagger(Y) \geq C, \nonumber \\
    &\quad Y = Y^\dagger,
\end{align}
with $\Xi^\dagger$ the dual map to $\Xi$. The exact optimization sense is unimportant, since it can be changed from minimisation to maximisation and vice-versa by changing the sign of the objective function. The exact prescription for $\Xi$, $C$ and $D$ for the diamond norm can be found in \cite{Watrous_NormSDP_2012}. In the next subsection, the more relevant prescription for the problem introduced in Eq. \eqref{SDP_CPTP_Div_Norm} will be provided.

\subsection{Application to the divisibility problem}

Combining the SDP above for the diamond norm with Eq.\ \eqref{SDP_CPTP_Div_Norm} we obtain the full SDP formulation for our divisibility problem:
\begin{align} \label{eq:app_sdp_cptp_norm}
    \text{Given}&\quad \varrho_{2|0},\, \varrho_{1|0} \nonumber \\
    \Minimise_{\varrho_{2|1},\, u_0,\, u_1,\, Y_0,\, Y_1} &\quad \frac{1}{2}(u_0 + u_1) \nonumber \\
    \textrm{s.t.} &\quad u_0\,\mathbb{1} \geq \mathrm{Tr}_{\mathcal{H}_{2}}[Y_0], \nonumber \\
    &\quad u_1\,\mathbb{1} \geq \mathrm{Tr}_{\mathcal{H}_{2}}[Y_1], \nonumber \\
    &\quad \begin{bmatrix}
    Y_0 & -\mathcal{C}_\Phi \cr
    -\mathcal{C}^\dag_\Phi & Y_1
    \end{bmatrix} \geq 0, \nonumber \\
    & \quad \mathcal{C}_{\Phi} = \varrho_{2|0} - \mbox{Tr}_{\mathcal{H}_{1}}\left[(\mathds{1}_{\mathcal{H}_{0}} \otimes \varrho_{2|1})^{T_{\mathcal{H}_{1}}} (\varrho_{1|0} \otimes \mathds{1}_{\mathcal{H}_{2}}) \right], \nonumber \\
    & \quad \varrho_{2|1} \geq 0, \nonumber \\
    & \quad \mbox{Tr}_{\mathcal{H}_{2}}\left[ \varrho_{2|1} \right] = \mathds{1}_{\mathcal{H}_{1}}.
\end{align}

It should be noted that, although one of the optimization variables, $\varrho_{2|1}$, appears in an expression containing tensor products and matrix products with other quantities, these are constant within the optimization and the constraint is still linear. To put it more explicitly, the optimization in Eq.\ \eqref{eq:app_sdp_cptp_norm} can be thought as the dual formulation of the problem. In the standard form of Eq.\ \eqref{eq:app_std_form_dual}, the variable $Y$ should be understood as containing the relevant variables of the problem in a block structure as follows:
\begin{equation}
    Y\,=\,\begin{bmatrix}
        u_0 & \cdot & \cdot & \cdot & \cdot \cr 
        \cdot & u_1 & \cdot & \cdot & \cdot \cr
        \cdot & \cdot & Y_0 & \cdot & \cdot \cr
        \cdot & \cdot & \cdot & Y_1 & \cdot \cr
        \cdot & \cdot & \cdot & \cdot & \varrho_{2|1}
    \end{bmatrix}.
\end{equation}
All other entries are free, but since they will not be relevant to the problem, they can be assumed to be equal to zero. From the block structure of $Y$, variables can be treated in an independent form. For instance, the coefficients of the objective function, $D$, are given simply by
\begin{equation}
    D\,=\,\frac{1}{2}\,\mathbb{1}_2 \oplus \boldsymbol{0},
\end{equation}
where $\mathbb{1}_2$ is the identity in $\mathbb{C}^2$ and $\boldsymbol{0}$ the null matrix acting on the remaining spaces, combined block diagonally.

For $\Xi^\dagger$, it is possible to combine independent channels on the variables by pinching the relevant variables with the adequate projectors and combining the constraints in a block-matrix structure. Equalities in this formulation can be achieved by combining two inequalities as $\textrm{lhs} \geq \textrm{rhs}$ and $-\textrm{lhs} \geq -\textrm{rhs}$ where $\textrm{lhs}$ and $\textrm{rhs}$ are the left-hand side and right-hand side of the constraint, respectively. Eventual constant terms that appear isolated from the optimization variables are left for the matrix $C$, in the corresponding entries.

To achieve a constraint such as 
\begin{equation}
    \begin{bmatrix}
    Y_0 & -\mathcal{C}_\Phi \cr
    -\mathcal{C}^\dag_\Phi & Y_1
    \end{bmatrix} \geq 0,
\end{equation}
with $\mathcal{C}_{\Phi} = \varrho_{2|0} - \mbox{Tr}_{\mathcal{H}_{1}}\left[(\mathds{1}_{\mathcal{H}_{0}} \otimes \varrho_{2|1})^{T_{\mathcal{H}_{1}}} (\varrho_{1|0} \otimes \mathds{1}_{\mathcal{H}_{2}}) \right]$, the part containing the variable $\varrho_{2|1}$ can be included in the l.h.s. of the inequality, while $\varrho_{2|0}$ is left as the corresponding entry in $C$. Then the trace and products in $\mathcal{C}_{\Phi}$ are expanded in terms of the entries of $\varrho_{2|1}$. For instance, the element $|\alpha,\beta\rangle\langle \alpha',\beta'|$ of the partial trace is given by
\begin{equation}
    \sum_{\kappa,\lambda} \left(\varrho_{1|0}\right)_{\alpha \kappa}^{\quad\alpha' \lambda}\,\left(\varrho_{2|1}\right)_{\lambda \beta'}^{\quad \kappa \beta},
\end{equation}
where a simplified notation has been used for the indices, combining the subspaces that compose the operators, such that $\varrho_{1|0} = \sum_{u,v,u',v'} (\varrho_{1|0})_{uv}^{\quad u'v'}\,|u,\,v\rangle\langle u',\,v'|$, and similarly for $\varrho_{2|1}$. It should be noted that the term is completely linear in $\varrho_{2|1}$, since $\varrho_{1|0}$ is given as a parameter to the problem, and $\Xi^\dagger$ projects $Y$ in its relevant terms, then adds an offset in the larger resulting constraint matrix, so that it acts as an independent constraint from the others.

It should be remarked that, since the SDP admits a primal and a dual formulation [Eqs. \eqref{eq:app_std_form_primal} and \eqref{eq:app_std_form_dual}] and, since it is the case that the values of both optimizations coincide \cite{Watrous2018}, an alternative formulation of problem \eqref{SDP_CPTP_Div_Norm} is possible and which results in an object that might be used as a witness for non-divisibility for a fixed $\varrho_{1|0}$,
\begin{align} \label{eq:app_sdp_witness}
    \text{Given}&\quad \varrho_{2|0},\,\varrho_{1|0} \nonumber \\
    \Maximise_{W,\, \Psi_0,\, \Psi_1,\, \Lambda} &\quad \mathrm{Tr}\left[W\,\varrho_{2|0}\right] + \mathrm{Tr}\left[\Lambda\right] \nonumber \\
    \textrm{s.t.} &\quad \begin{bmatrix}
    \Psi_0 \otimes \mathbb{1}_{\mathcal{H}_{2}} & W \cr
    W^\dagger & \Psi_1 \otimes \mathbb{1}_{\mathcal{H}_{2}}
    \end{bmatrix} \geq 0, \nonumber \\
    &\quad \mathrm{Tr}[\Psi_0] = 1,\,\,\mathrm{Tr}[\Psi_1]=1, \nonumber \\
    &\quad \Lambda \otimes \mathbb{1}_{\mathcal{H}_{2}} \leq -\mathrm{Tr}_{\mathcal{H}_{0}}\left[((\varrho_{1|0})^{T_{\mathcal{H}_{1}}} \otimes \mathbb{1}_{\mathcal{H}_{2}} )(W \otimes \mathbb{1}_{\mathcal{H}_{1}})\right].
\end{align}

Assume that $\varrho_{2|0}$ is CP-divisible with factors $\varrho_{2|1}$ and $\varrho_{1|0}$. Since, by assumption, $\varrho_{2|1}$ is positive-semidefinite, let $K = \mathbb{1}_{\mathcal{H}_{0}} \otimes \sqrt{\varrho^{\vphantom{'}}_{2|1}}$. The last constraint implies that
\begin{equation}
    K \left(\Lambda \otimes \mathbb{1}_{\mathcal{H}_{2}}\right) K \leq -\mathrm{Tr}_{\mathcal{H}_{0}}\left[K\,((\varrho_{1|0})^{T_{\mathcal{H}_{1}}} \otimes \mathbb{1}_{\mathcal{H}_{2}} )(W \otimes \mathbb{1}_{\mathcal{H}_{1}} )\,K\right],
\end{equation}
which, in turn, implies that
\begin{align}
    \mathrm{Tr}&\left[\left(\Lambda \otimes \mathbb{1}_{\mathcal{H}_{2}}\right)\,(\mathbb{1}_{\mathcal{H}_{0}} \otimes \varrho_{2|1})\right] \leq \nonumber \\
    &-\mathrm{Tr}\left[(\mathbb{1}_{\mathcal{H}_{0}} \otimes \varrho_{2|1})\,((\varrho_{1|0})^{T_{\mathcal{H}_{1}}} \otimes \mathbb{1}_{\mathcal{H}_{2}} )(W \otimes \mathbb{1}_{\mathcal{H}_{1}})\right],
\end{align}
since $\mathcal{O} \geq 0 \,\Rightarrow\,\mathrm{Tr}[\mathcal{O}] \geq 0$.

Then, noting that, due to trace preservation, $\mathrm{Tr}_{\mathcal{H}_{2}}[\varrho_{2|1}] = \mathbb{1}_{\mathcal{H}_{1}}$, we conclude that the left-hand side of the inequality above, $\mathrm{Tr}\left[\left(\Lambda \otimes \mathbb{1}_{\mathcal{H}_{2}}\right)\,(\mathbb{1}_{\mathcal{H}_{0}} \otimes \varrho_{2|1})\right]$, coincides with $\mathrm{Tr}[\Lambda]$. Finally then, using this bound on the trace of $\Lambda$, it is possible to conclude that the optimal value of Eq. \eqref{eq:app_sdp_witness} is bounded above by
\begin{equation}
   \mathrm{Tr}\left[W\left(\varrho_{2|0} - \mathrm{Tr}_{\mathcal{H}_{1}}[(\mathbb{1}_{\mathcal{H}_{0}} \otimes \varrho_{2|1})\,((\varrho_{1|0})^{T_{\mathcal{H}_{1}}} \otimes \mathbb{1}_{\mathcal{H}_{2}})]\right)\right] = 0,
\end{equation}
for a CP-divisible $\varrho_{2|0}$. This proves that the pair $(W,\,\Lambda)$ serves as a witness for divisibility for a fixed $\varrho_{1|0}$, with the necessary condition for divisibility 
\begin{equation}
\mathrm{Tr}[W\,\varrho_{2|0}] + \mathrm{Tr}[\Lambda] \leq 0,
\label{eq:app_dual_condition}
\end{equation}
whenever $\{\varrho_{2|0},\,\varrho_{1|0}\}$ forms a CP-divisible dynamics, but potentially having a positive value when it is not.

\section{A notion of absolute non-divisibility}
\label{App.AbsoluteNondiv}

In both applications presented in the main text, we notice that, for particular values of the parameters, the optimal channels returned by the SDPs are the identity channel, meaning that any tentative channel combined with $\Lambda_{1|0}$ as a factor to approximate $\Lambda_{2|0}$ tends only to make them more dissimilar, as measured by the diamond norm. Motivated by this finding, below we introduce and motivate a notion of absolute non-divisibility.

Notice that for any given CPTP maps $\Psi_0,\,\Psi_1,\,\Phi_0,\,\Phi_1$, the diamond norm satisfies the property
\begin{equation}
    \| \Psi_1 \circ \Psi_0 - \Phi_1 \circ \Phi_0\|_\diamond \leq \| \Psi_1 - \Phi_1 \|_\diamond + \| \Psi_0 - \Phi_0\|_\diamond,
\end{equation} 
i.e. errors in performing a composition of maps are bounded additively on the individual errors of each map applied. To apply this property to the divisibility task, assume that the target map $\Lambda_{2|0}$ can be split generically into two factors $\Lambda_B$ and $\Lambda_A$, in principle unrelated to the problem, but such that
\begin{equation}
    \Lambda_{2|0} = \Lambda_B \circ \Lambda_A.
\end{equation}

Choosing a particular decomposition for $\Lambda_{2|0}$, 
we obtain
\begin{equation}
    \| \Lambda_{2|0} - \Lambda \circ \Lambda_{1|0} \|_\diamond = \| \Lambda_B \circ \Lambda_A - \Lambda \circ \Lambda_{1|0} \|_\diamond,
\end{equation}
and boundedness of the norm implies 
\begin{equation}
    \| \Lambda_{2|0} - \Lambda \circ \Lambda_{1|0} \|_\diamond \leq \dnorm{\Lambda_B - \Lambda} + \dnorm{\Lambda_A - \Lambda_{1|0}},
\end{equation}
for any candidate map $\Lambda$ that should be composed with $\Lambda_{1|0}$ in the attempt to obtain exact divisibility. 

Given the factors $\Lambda_B$ and $\Lambda_A$, an optimal choice for $\Lambda$ to tighten the bound above is $\Lambda = \Lambda_B$. With this choice, the upper bound reduces to $\dnorm{\Lambda_A - \Lambda_{1|0}}$, which measures how well $\Lambda_{1|0}$ can be approximated by the factor $\Lambda_A$. Indeed, it is intuitive to conjecture that the problem of divisibility can be equivalently restated as finding the closest factor $\Lambda_A$ to the starting channel $\Lambda_{1|0}$ (with optimal divisibility attained when they are equal), then taking $\Lambda_{1|0}$ as a noisy version of $\Lambda_A$, so that composition with $\Lambda_B$ results in the best strategy to approximate $\Lambda_{2|0}$. 

Two possible trivial options for splitting $\Lambda_{2|0}$ are available in every case: 
(i) $\Lambda_A = \Lambda_{2|0}$, $\Lambda_B = \mathcal{I}$; (ii) $\Lambda_A = \mathcal{I}$, $\Lambda_B = \Lambda_{2|0}$. In the former, optimal $\Lambda$ is given by the identity channel, meaning that nothing can be done on $\Lambda_{1|0}$ to improve on its closeness towards $\Lambda_{2|0}$. In the latter case, $\Lambda_{1|0}$ is so close to the identity channel that the best strategy is to use $\Lambda = \Lambda_{2|0}$ itself and consider the error imposed by $\Lambda_{1|0}$ as a perturbation on exact identity.
Thus, trivially, we obtain the two upper bounds to optimal divisibility:
\begin{equation}
\dnorm{\Lambda_{2|0} - \Lambda \circ \Lambda_{1|0}} \leq \min\{ \dnorm{\Lambda_{2|0} - \Lambda_{1|0}},\, \dnorm{\mathcal{I} - \Lambda_{1|0}} \}.
\end{equation}

We may call \emph{absolute non-divisibility (with respect to the diamond norm)} between maps $\Lambda_{2|0}$ and $\Lambda_{1|0}$ whenever they are not trivial with respect to the bound above, i.e. when $\Lambda_{2|0} \neq \Lambda_{1|0}$ and $\Lambda_{1|0} \neq \mathcal{I}$, and yet the result of the semidefinite optimization is either $\Lambda = \mathcal{I}$ or $\Lambda = \Lambda_{2|0}$, meaning that optimal solution is the bound for trivial decomposition for a non-trivial couple of maps $\Lambda_{2|0}$ and $\Lambda_{1|0}$. 

More generally, a trivial splitting of $\Lambda_{2|0}$ could involve an input or output unitary transformation, so that factorization of $\Lambda_{2|0}$ would be given by $\Lambda_A = \mathcal{U}^{-1} \circ \Lambda_{2|0}$, $\Lambda_B = \mathcal{U}$, where $\mathcal{U}$ is a unitary channel, or $\Lambda_B = \Lambda_{2|0} \circ \mathcal{U}^{-1}$ and $\Lambda_A = \mathcal{U}$. The same analysis as before follows, with $\mathcal{U} = \mathcal{I}$ being a particular case: either $\Lambda_{1|0}$ is too close to $\mathcal{U}^{-1}\circ \Lambda_{2|0}$ to obtain any improvement with a different factor, or $\Lambda_{1|0}$ is too close to the unitary $\mathcal{U}$ so that approximating $\Lambda_{2|0}$ just entails correcting the input basis and treating $\Lambda_{1|0}$ as a perturbation.

\bibliography{list_of_references}
\end{document}